\documentclass[twocolumn,aps,amsmath,amssymb,prb,showpacs,superscriptaddress]{revtex4-1}

\usepackage{graphicx}
\usepackage{amsmath}
\usepackage{amssymb}
\usepackage{bm}
\usepackage{color}
\usepackage{amsfonts}
\usepackage{appendix}
\usepackage{color}
\usepackage{dsfont}
\usepackage{soul}
\setstcolor{red}

\definecolor{lightgrey}{rgb}{0.7,0.7,0.7}

\newcommand{\sgn}{{\rm sgn}}

\newcommand{\ignore}[1]{\relax}

\begin{document}
\title{
Intermediate fixed point in a Luttinger liquid with elastic and dissipative backscattering
}
\author{Alexander Altland}
\affiliation{Institut f\"ur Theoretische Physik, Universit\"at zu K\"oln, K\"oln, 50937, Germany.}
\author{Yuval Gefen}
\affiliation{Department of Condensed Matter Physics, The Weizmann Institute of Science,  Rehovot 76100, Israel.}
\author{Bernd Rosenow}
\affiliation{Institut f\"ur Theoretische Physik, Universit\"at Leipzig, D-04103, Leipzig, Germany.}

\date{March 19, 2015}
\begin{abstract}
In a recent work [Phys. Rev. Lett. {\bf 108}, 136401 (2012)]  we have addressed the problem of a Luttinger liquid with a scatterer that allows for both coherent and incoherent scattering channels. We have found that the physics associated with this model is qualitatively different from the elastic impurity setup analyzed by Kane and Fisher, and from the inelastic scattering scenario studied by Furusaki and Matveev, thus proposing a new paradigmatic picture of Luttinger liquid with an impurity. Here we present an extensive study of the renormalization group flows for this problem, the fixed point landscape, and scaling near those fixed points. Our analysis is non-perturbative in the elastic tunneling amplitudes, employing an instanton calculation in one or two of the available elastic tunneling channels. Our analysis accounts for non-trivial Klein factors, which represent anyonic or fermionic statistics. These Klein factors need to be taken into account due to the fact that higher order tunneling processes take place. In particular we find a stable fixed point, where an incoming current is split ${1 \over 2}$ - $1\over 2$ between a forward and a backward scattered beams. This intermediate fixed point, between complete backscattering and full forward scattering, is stable for the Luttinger parameter $g<1$.
\end{abstract}
\pacs{}

\maketitle

\section{Introduction}

The concept of a Luttinger liquid (LL) provides  a very general framework to deal with a strongly interacting electron gas confined to one spatial dimension (1D) \cite{Luttinger}.  Contrary to higher dimensional situations, where the quasi-particle (qp) concept describes low energy excitations, in a 
LL only collective excitations are long lived and stable. Particle-like excitations of a LL have an energy dependent density of states and can relax in energy in the presence of backscattering. Applications  of the LL concept include semiconductor  \cite{semiconductor }, metallic  \cite{metallic}, and polymer  \cite{polymer} nano wires, carbon nanotubes \cite{CNT}, quantum Hall edges \cite{QHedge}, and cold atoms \cite{cold_atoms}.

Due to the geometrical restriction to 1D,  electrons may be divided into two sectors, right moving and left moving. The presence of an impurity \cite{single_impurity}gives rise to inter-branch scattering, i.e.~backscattering with a power-law  dependence of the  scattering probability on energy. 
Hitherto there have been two paradigmatic models which addressed impurity scattering in the context of LL: a purely elastic impurity  as discussed by Kane and Fisher \cite{KF} (KF),  and totally inelastic scattering described by Furusaki and Matveev \cite{Matveev} (FM).  

In a recent publication \cite{AGR12}  we have introduced and studied a model which is a hybrid between the two. We have addressed interacting  electrons in one dimension described by a Luttinger model, which includes a scatterer that  may give rise to elastic coherent tunneling, and at the same time may accommodate inelastic modes. Our analysis has led to predictions which are qualitatively different from the KF and the FM pictures, and in this sense can be considered as a new paradigmatic  scheme of impurity scattering in the context of LL.

In particular,  we have obtained  the following results:
i) Due to the presence of both elastic and inelastic scattering channels, there exists a new stable    non Fermi liquid fixed point (FP). Asymptotically, the impurity becomes a symmetric beam splitter, which halves the incoming beam into two equal outgoing beams. For this reason we termed this limit a  ${1 \over 2}$ - $1\over 2$   FP. We also identify other non Fermi liquid FPs, which are stable in a certain direction and unstable in another direction in parameter space. A major facet  of our earlier work was that upon renormalizing our model down in temperature (or bias voltage), under generic conditions the model flows to a stable FP. The neighborhood of this  FP is marked by the non-Fermi liquid correlations described above.
ii) At equilibrium, there is thermal noise at  each incoming and  each outgoing terminal, but no cross-correlation in the thermal noise, i.e.~no correlations between incoming-incoming and outgoing-outgoing edges - in stark contrast to  the Landauer-Buttiker picture. 
iii) Out of equilibrium, when the system is voltage biased at one of the incoming terminals, the impurity acts as a ${1 \over 2}$ - $1\over 2$  beam splitter with   no shot noise component in the outgoing current. Similarly,  cross-current correlators do not contain a shot noise component.

Our analysis in reference [\onlinecite{AGR12}] involved bosonization of the interacting fermionic Hamiltonian. It was based on perturbation in the elastic (coherent) scattering amplitudes in the presence of a charging term representing electrostatic correlations in the scattering region.  We have concluded that for energies below the charging energy, the bosonic fields affected by the elastic scattering terms  become massive. As a result,  only one of the four independent bosonic fields in the model remains asymptotically free. The asysmptotically Gaussian action has facilitated calculation of noise and current-current correlations. 

Naturally, the fact that our analysis has been based on, and the results have been motivated by perturbative analysis, raises some questions. Most important is the fact that various relevant terms of the action have been analyzed within a renormalization group (RG) scheme separately. Evidently this procedure ceases to be justified when the respective amplitudes of the elastic tunneling terms   grow  through the RG analysis. One may need to evaluate expectation values of products of non-commuting terms, which a naive perturbative analysis is incapable of doing. Careful analysis is required in this process. 

We note that the elastic  tunneling channels involve scattering from any of the two incoming channels to any of the outgoing channels. To establish the right language, and to enable a compact description of the RG flow diagram, we divide the elastic scattering terms into Òback scatteringÓ (the terms connecting channels 1 with 3, and 2 with 4, cf. Fig. 2) and Òforward scatteringÓ (the terms connecting 1 with 4, and 2 with 3). We assume that the coefficients of the two back scattering terms are equal to each other, and similarly the coefficients of the two forward scattering are equal to each other. Relaxing this  assumption does not modify the asymptotics of the problem, but makes the description of the RG flows more cumbersome.  Assuming we have renormalized the problem down from high temperature (or voltage), and that certain irrelevant non-universal  terms have been renormalized out, we are left with flows in a two-dimensional parameter space. We indeed establish the existence of a new stable  ${1 \over 2}$ - $1\over 2$ FP, which corresponds to strong coupling in both scattering channels (backscattering and forward-scattering respectively). We denote this a  strong-strong fixed point   (SSFP) . In addition,  we find and characterize three more FPs: a trivial unstable weak scattering FP (weak-weak fixed point,  WWFP), and two other FPs:   a weak-strong fixed point (WSFP),  and a strong-weak fixed point (SWFP), each being stable in one direction of  parameter space and unstable in the other direction. All these points are marked by  non Fermi liquid correlations. 

We note that the issue of non-commutativity of the various  tunneling operators is central to  the analysis of the model beyond the perturbative limit. Here,  we will be dealing with operators describing tunneling of Abelian quasi-particles. Such particles (known as anyons) possess fractional statistics, intermediate between bosons and fermions. The exchange of two qps involves a statistical phase which is ill-defined: its sign depends on the "history" of the trajectories employed in the course of the exchange. In order to avoid such a pathology, one may resort to either one of the three following tricks: i) avoid single qp field operators (qp creation or annihilation operators) and assign  statistical Klein factors only to operators which are bilinear in qp field operators, e.g. to qp tunneling operators \cite{Kane2003,Law+06}, ii) attach a statistical flux tube to a qp and follow the kinetics by way of a quantum  Master equation \cite{Feldman+07,Campagnano+12}, iii) introduce all relevant edges on a single contour, and choose a convention how qp's are being exchanged \cite{Chamon} (i.e., clockwise or anticlockwise). In the present analysis we will  assign generalized Klein factors to qp operators, following the philosophy of iii). 

The outline of the paper is the following: 
In Section II,  we introduce the problem addressed here, and discuss its mapping onto a model setup which allows  systematic bosonization and subsequent  analysis. The assumptions underlining this modelization and the applicability thereof are outlined. In Section III,  we position our building blocks (in particular the  chiral  channels of our model) within  a particular geometry, giving rise to a well-defined convention that determines the commutation relations of the anyonic quesi-particles. Performing  a sequence of canonical transformations allows us to derive an effective action that captures the symmetries of the problem.  Section IV is devoted to the perturbative analysis of the weak scattering fixed point (WWFP), while  Section V is focused on  the study of the scaling near the weak-strong fixed point WSFP. We note that the physics of the strong-weak  fixed point SWFP is trivially obtained by trivial exchange of field indices from the WSFP.   Here,  we argue that the low energy dynamics is dominated by a  phase slip-instanton mechanism. In Section VI,  we analyze the scaling near the SSFP.  In this limit,  an instanton picture within a 2-dimenional parameter space is employed. In the limit of non-interacting Luttinger wires our analysis needs special care: some of the flows become  marginal. This is discussed in Section VII, where comparison with known results is presented as a consistency check.   Finally, in Section VIII, we present a short summary of our results, and a few proposals and speculations concerning further directions of our analysis. In Appendix A,  we make a few comments concerning experimental verification of our predictions, and in Appendix B, we compare the results obtained within our bosonic theory with an exact refermionization treatment for a specific value of the Luttinger parameter.

\section{The setup and its modeling}

\begin{figure}[h]
\includegraphics[width=0.8\linewidth]{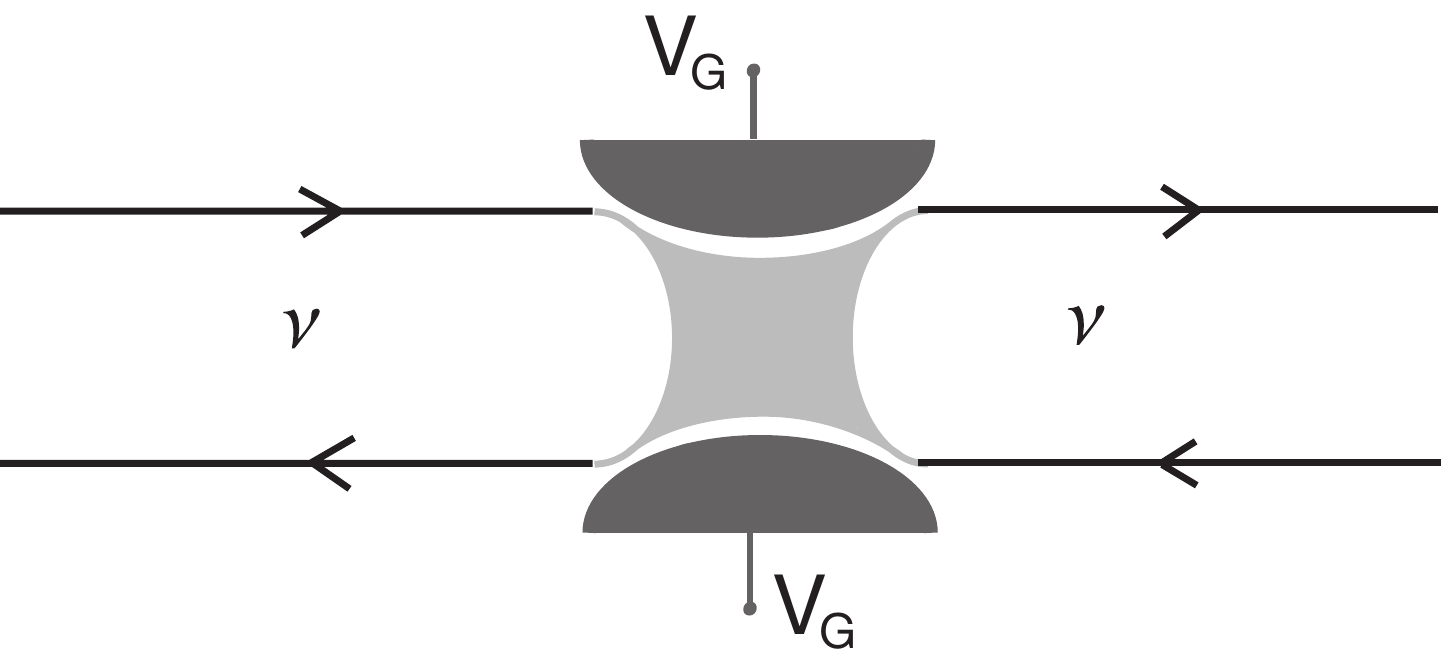}
\caption{Sketch of a QH bar with a gate defined QPC. The bulk of the Hall bar 
is incompressible with filling fraction $\nu$, the density in the QPC region is
lower, such that mixing between edge states in a compressible region can occur, both via elastic and inelastic channels.   }
\label{fig:1}
\end{figure}

Our model is motivated by an experimental setup with a two-dimensional electron gas in a strong magnetic field, such that the 
fractional quantum Hall (FQH) regime is reached. We focus on a  Laughlin filling fraction, such that, 
in the absence of edge reconstruction, the structure of the edge is simple, consisting of a single chiral  channel at each edge. 
In addition, we imagine the existence of a compressible puddle with somewhat reduced density,  e.g.~$\nu=1/3$ in the bulk and a puddle with a compressible   $\nu=1/4$ state of composite fermions. Such a setup can be realized  by  locally modifying the backgate voltage in a certain region, see Fig.~\ref{fig:1}. 
%
%

%
\begin{figure}
\includegraphics[width=8cm]{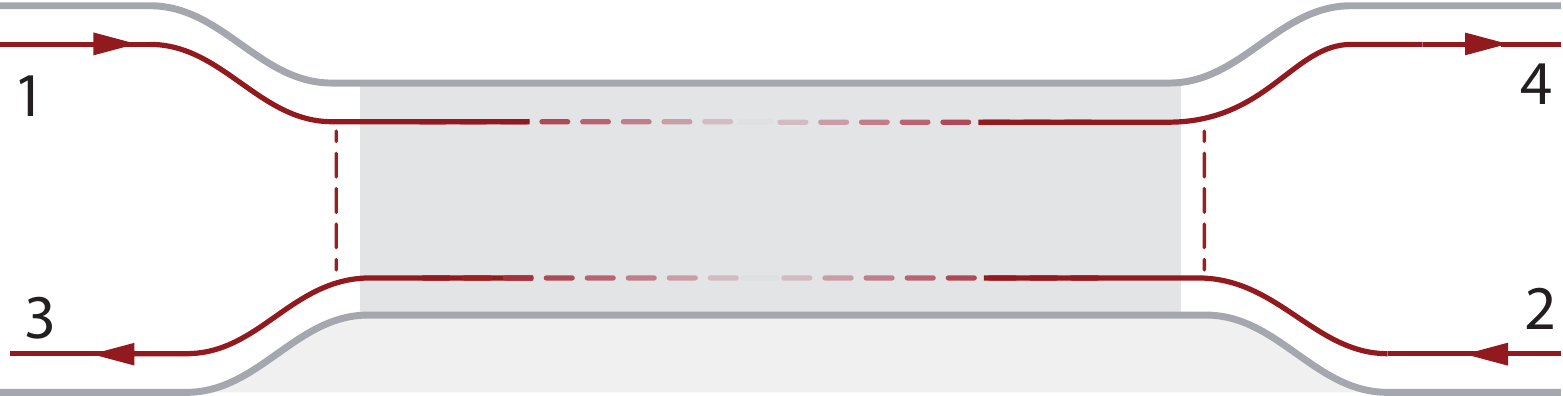}
\caption[]{
(Color online) Schematic model of a quantum wire with two incoming (labeled 1,2) and two
outgoing  chiral
modes (labeled 3,4) connecting to an extended scattering region. 
The dark shading represents a region of
capacitive charging, the quantum dot.
Near the entrance and exit to the quantum dot, where
translational invariance is broken, backscattering of quasi-particles across the Hall bar is possible, 
denoted by vertical dashed lines.  In a possible realization with 
a compressible quantum dot region, the edge current spreads into the bulk, and 
both elastic transmission along the edge (indicated by horizontal dashed line) and inelastic transmission through the dot is possible.}
\label{fig:setup}
\end{figure}
%

  Due to its small geometric size, the puddle has a charging energy associated with it and constitutes a quantum dot (QD). 
The edge modes support 
anyonic qps, which may  be backscattered through the incompressible bulk near the entrance and exit of the QD. 
The setup we thus envision is that of a QD with four semi-infinite chiral edges coupled to it. We  assume that the contacts between the leads and the QD are ideal, i.e.~reflectionless. In total, one needs to  consider a Hamiltonian which describes the ideal chiral  edges, the contact between edges and puddle, the puddle modeled as a metallic quantum dot (QD) with charging energy,   and elastic (coherent) tunneling channels  that may support qp tunneling (see  Fig.~2). 
In terms of modeling this setup, we need to find a way to  describe an ideal, reflectionless contact between a chiral lead and a QD. In order to achieve this goal, we follow Matveev: we employ four electrostatically coupled infinite chirals. For each of these infinite chirals, one half is identified with the chiral wire outside the dot, whereas the other half is assumed to be positioned inside the dot. In this way, low energy excitations of the chiral parts "inside the dot" can be identified with dissipative processes within the QD. Transport processes between a chiral entering the dot and another chiral leaving the dot are mediated by i) the charging energy within the dot, which prevents charge accumulation inside the QD, and ii) by coherent elastic tunneling between incoming and outgoing chirals.  The setup corresponding to our model is depicted  in Fig.~3. 

We are now in a position to write the total  Hamiltonian . It consists of three terms: The low energy dynamics of the original semi-infinite leads, as well as the part of the chiral edge channels that constitutes partof the quantum dot, 
are delegated ot a Hamiltonian that represents four infinite chirals, $ H_{\rm ch}$. Separately, the charging energy of the QD is described by $ H_{\rm dot}$. Finally, the coherent part of the lead-QD tunneling is represented
by $H_{\rm tun}$. We resort to bosonization, and  define  bosonic fields $\phi_i(x)$ with $i= 1, 2, 3, 4$ for each chrial, see Fig.~2. We 
decompose the bosonic fields into finite momentum and zero modes according to 
%
\begin{equation}
\varphi_i(x) = \phi_i(x)  + \phi^0_i(x)   \ \ .
\end{equation}
%
The finite momentum part has a Fourier representation according to 
%
\begin{equation}
\phi_i(x) \ = \ \int {d q \over 2 \pi}   \theta(\pm q)   \sqrt{2 \pi \nu \over |q|} \left( a_{i,q} e^{i q x} + a^\dagger_{i,q} e^{- i q x} \right)  \ ,
\end{equation}
%
and the structure of the zero modes will be discussed in the context of Klein factors in Section III.  For the mode decomposition above,  
 the plus sign should be used for the right-moving modes 1 and 4, and the minus sign for the left moving modes 2 and 3. The $a_{i,q}$ and 
 $a^\dagger_{i,q}$ are canonical boson operators with commutation relations $[ a_{i,q}, a^\dagger_{j,q^\prime}] = \delta_{i,j} 2 \pi 
 \delta(q - q^\prime)$, giving rise to $[\varphi_i(x),\varphi_j(y) ] = (\pi g /2) \sgn(x -y)$. The charge density is related to be boson fields via $\rho_i(x) = {1 \over \sqrt{2} \pi} \partial_x \phi_i(x)$, and the Hamiltonian for the chirals is given by
 %
 \begin{equation}
 H_{\rm ch} =   {\hbar v \over 2 \pi g}  \sum_{i=1}^4    \int dx \left[ \partial_x \varphi_i(x) \right]^2 \ \ \ .
\end{equation}
%
 In addition, there is a Coulomb charging energy for the dot
 %
 \begin{equation}
 H_{\rm dot} = {e^2 \over 2 C}  Q^2 \ \ , 
 \end{equation}
%
where 
%
 \begin{equation}
 Q = {1 \over \sqrt{2} \pi} \left[ - \phi_1(0) - \phi_2(0) + \phi_3(0) + \phi_4(0) \right]
 \end{equation}
%
denotes the charge on the dot. Coherent elastic tunneling between incoming and outgoing wires is described by 
the  Hamiltonian
%
\begin{eqnarray}
\label{ScatteringH}
H_{\rm tun}  & = & t_{13} \cos\left( \varphi_1 - \varphi_3 \right) + t_{24} \cos\left( \varphi_2 - \varphi_4 \right) \\
& & + t_{14} \cos \left( \varphi_1 - \varphi_4\right)  + t_{23} \cos \left( \varphi_2 - \varphi_3 \right) \nonumber \ \ . 
\label{Htun.eq}
\end{eqnarray}
%
Here we ignore retardation effects, assuming that tuneling is instantaneous. Thus, the total Hamiltonian for the system is given by
%
\begin{equation}
H = H_{\rm ch} + H_{\rm dot} + H_{\rm tun}  \ \ .
\end{equation}
%
As for some parts of our analysis it will be useful to resort to a functional integral formulation of the problem, 
we also state the Matsubara finite temperature action of the system 
%
\begin{equation}
S = { \hbar  \over 2 \pi g} \sum_{j=1}^4      \int   \! \!      dx  \, d \tau \,  \partial_x \varphi_j(x,\tau)  \left[\pm i \partial_\tau \varphi_j(x,\tau)\right]  
- \int d \tau H   \ \ . 
\end{equation}
%
Here, the plus sign refers to the right moving branches 1 and 4, and the minus sign to the left moving branches 2 and 3. 
We note that the zero mode part of the Hamiltonian $H_0$ and and of the action $S_0$  will be discussed in detail in Section III. 

%
\begin{figure}
\includegraphics[width=7cm]{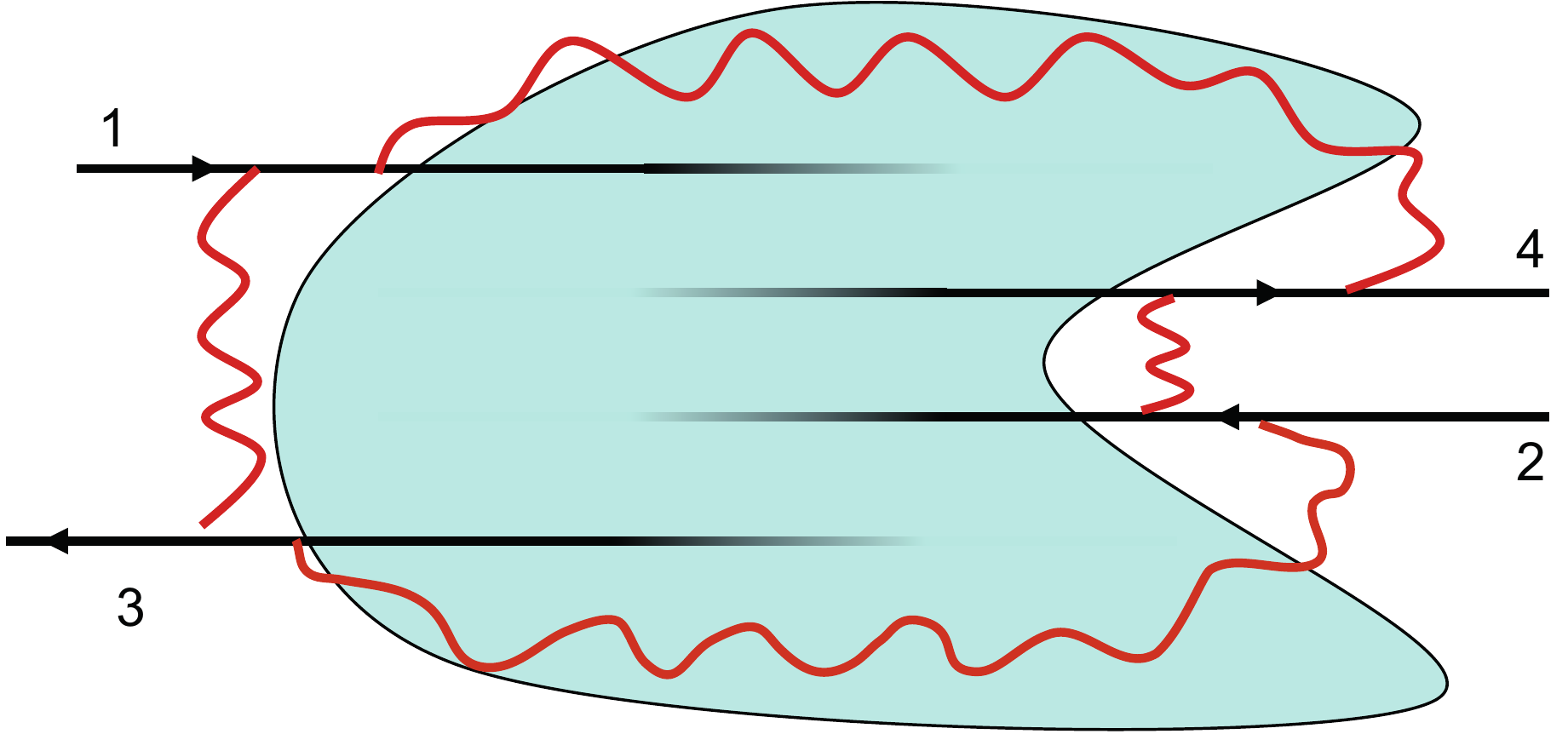}
\caption[]{
(Color online)   Effective modelling of the system (cf. Fig.~\ref{fig:setup}) through four infinite chiral channels. Each channel consists of a half infinite wire (either incoming or outgoing), and a half infinite wire which is part of the quantum dot (shaded area). Elastic scattering channels are represented by wiggly lines. }
\label{fig:fig3effective}
\end{figure}
%

 A few commons are in order now: (i)  Our analysis will be performed  using
 the Matsubara technique. (ii)  The tunneling is that of anyons (qp�s). We
 consider {\em only}  this type of tunneling terms as they are (when allowed)
 potentially the most relevant terms. (iii) All channels are put on an equal
 footing. There is no  distinction between �forward scattering � and
 �backscattering� any more. Particles on different chirals are  different
 species, hence the finite momentum boson fields  $\phi_1$, $\phi_2,...$
 commute. It is therefore vital (for the sake of preserving the appropriate
 commutation relations) to introduce Klein factors which, in the case of qp
 tunneling, will be anyonic Klein factors. In the notation of
 Eq.~\eqref{ScatteringH}, the Klein factors are implicit in the prefactors
 $t_{ij}$, their precise realization will be discussed below. We denote the $c$-number valued scattering amplitudes without Klein factors by $\gamma_{ij}$, i.e. $t_{ij}=\gamma_{ij}\times $ (Klein factors). (iv)  Each
 infinite chiral contributes half to the lead (either incoming  or outgoing)
 and half to the degrees of freedom of a compressible puddle.  The  puddle
 (a.k.a.~quantum dot) accommodates gapless soft modes. (v) The charging (that
 is, electrostatic)  interaction introduces coupling among the 4 channels.  As
 we renormalize the model down in temperature (or applied voltage), flowing
 towards low energy dynamics (long time scales),  the presence of a charging
 energy scale will force charge fluctuations on the QD to freeze out. The QD
 then remains effectively neutral. (vi) The model is a hybrid of elastic
 (coherent) scattering between incoming and outgoingleads, {\em
 and} inelastic transport through the QD. The latter involves the excitation
 of low-lying modes in the QD, which can be described in terms of many
 particle-hole excitations. The former represents (to lowest order in the
 tunneling coupling constant) elastic tunneling through the QD. While one may
 imagine that backward tunneling, say between channels 1 and 3, is mediated by
 a resonant impurity level in a Hall bar constriction, one can motivate the
 forward tunneling, say between channels 1 and 4, as being due to coherent
 propagation along the edge of the QD. For any real QD, there is an energy
 scale defined by the mean single-particle level spacing in the QD, $\Delta$.
 Once the maximum of temperature  and bias voltage is driven below $\Delta $,
 the soft modes which constitute an essential part of our model are frozen
 out. At that point our model ceases to be valid. Nevertheless, as long as the
 infrared cutoff of our RG flows is larger or comparable to $\Delta$ (which
 might be extremely small) the model and the  predictions following from its
 analysis, are valid. (vii) The interesting regime of our RG flows stipulates
 running frequencies $ \Delta \ll |\omega| \ll E_c$.  Modes involving charge
 fluctuations are frozen out; for the model to be valid,  soft modes in the QD
 still need to be available.   Is such a regime possible? If we are focusing
 on a QD subject to a strong perpendicular magnetic field,  then  the relevant
 spectrum of the QD is 1d (on the edge). Ostensibly, both $E_c$ and $\Delta$
 will scale like 1/L (L= the linear size of the QD), and the difference
 between them may be only due to different coefficients. However, we may
 control the  charging energy $E_c$, and actually render it larger than
 $\Delta$,  by considering a puddle realized by a sea of composite fermions,
 for instance at filling $\nu = 1/4$, whose motion does not follow 1d edges
 but instead explores the available 2d area of the QD.

Our analysis presented here is done with a view to the  chiral edges of a FQHE. We stress here that it may apply to quantum wires described by Luttinger liquid models. Applying our analysis to Luttinger liquids implies certain assumptions. This is discussed in Appendix A.

\section{  Generalized Klein factors, and the effective action}
  
As was commented in Section II, we need to establish a convention as far as Klein factors,  underlying the anyonic field operators, are concerned. We choose to establish such generalized Klein factors in relation to a geometry where all 4 infinite chirals are assumed to be segments of a single chiral contour. In that convention, the exchange of two anyons positioned on  channel  $i$   and  channel   $j$  respectively has an unambigious chirality. The geometry is shown in Fig.~4, and the related bosnonic field configuration is depicted in Fig.~5. 
%
\begin{figure}[h]
  \centering
  \includegraphics[width=7cm]{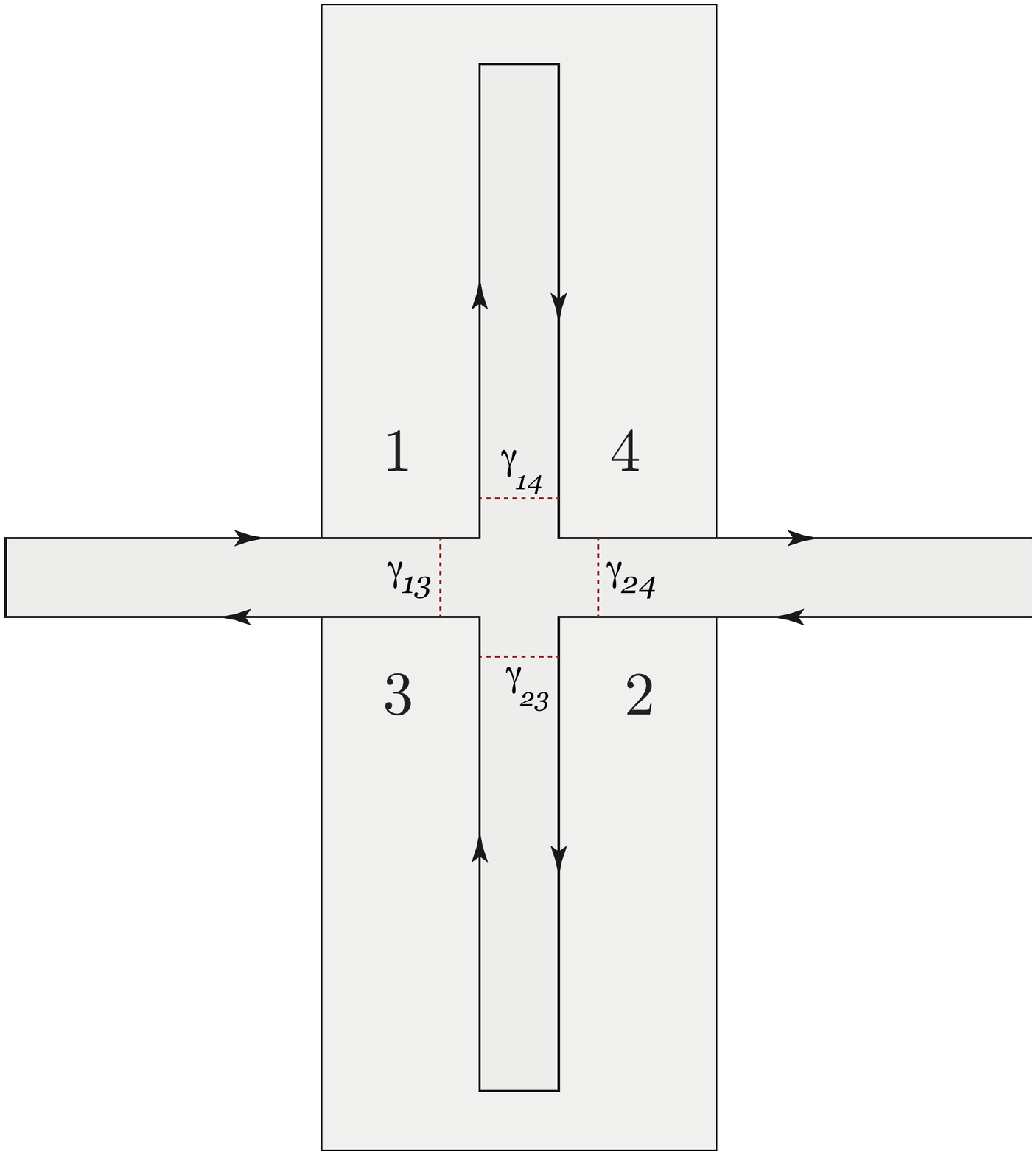}
  \caption{Putting all four chiral wires onto a single contour. Our
    previously disconnected chiral wires become segments denoted 1,2,3,4. They
    are connected by scattering as shown (cf. Eq.~(\ref{Htun.eq}). The quantum
    dot is represented by the shaded area. All segments should be imagined
    'infinitely long', e.g.~a quasi-particle entering the dot along 1 will then
    propagate vertically upwards but  will never make it to the 'top of the
    contour' in finite time. It has to scatter to get to 4 (henceforth to 2). 
    Note that the vertical scattering amplitude  $\gamma_{v}=|\gamma_{13} + e^{- i \pi g} \, \gamma_{24}|$
    and the horizontal scattering amplitude
 $\gamma_h=|\gamma_{23} + \gamma_{14}|$ 
 will be the parameters of our RG flow as shown in Fig.~\ref{rgflow.fig}. The relative phase factor in the 
 definition of $\gamma_v$ is due to the phases of scattering operators in Eq.~(\ref{eq:39}).}
  \label{fig:ChiralContour}
\end{figure}
%

 %
\begin{figure}[h]
  \centering
  \includegraphics[width=8cm]{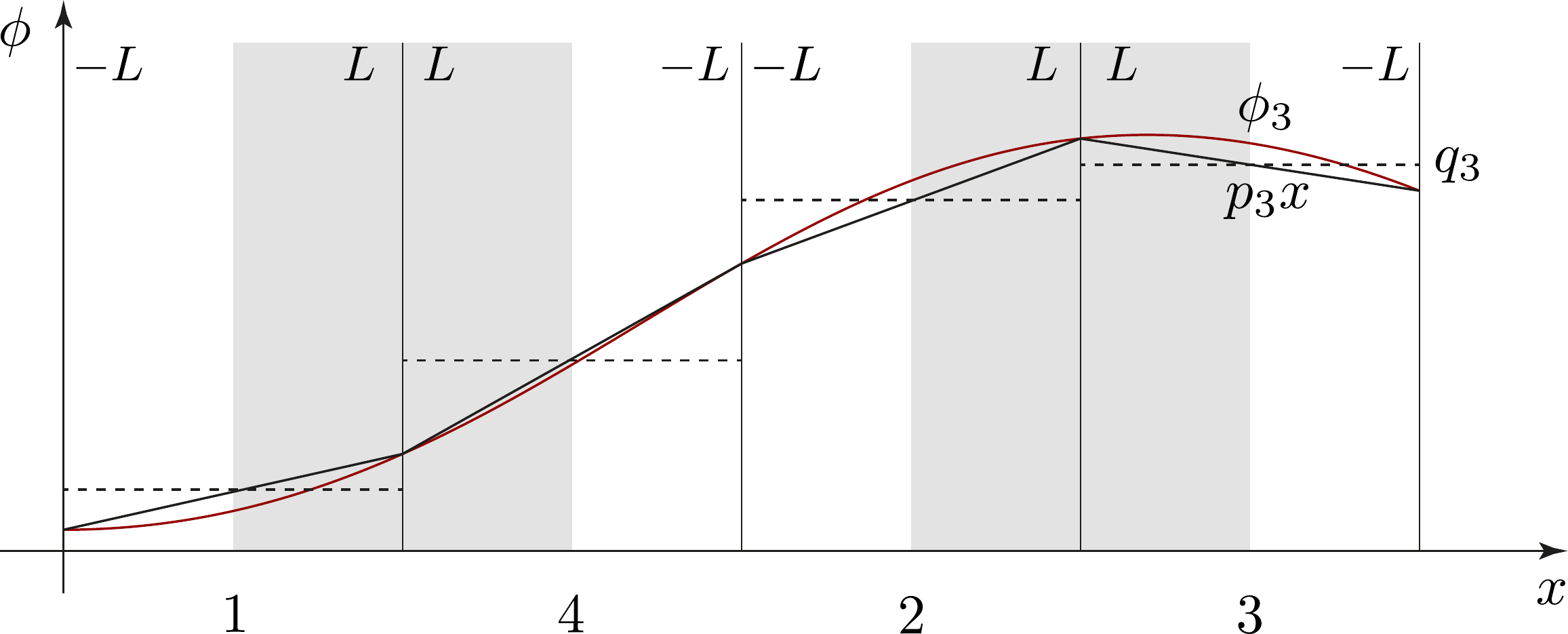}
  \caption{Field configuration on the closed contour. We introduce
    zero modes $q_i$, $p_i$ to describe the increment of the field
    throughout any of the four segments. The 'finite momentum
    components' $\phi_i$ obey periodic boundary conditions. Shaded
    regions indicate the dot regions. }
  \label{fig:ZeroModes}
\end{figure}
%

%
\begin{figure}[h]
\includegraphics[width=0.8\linewidth]{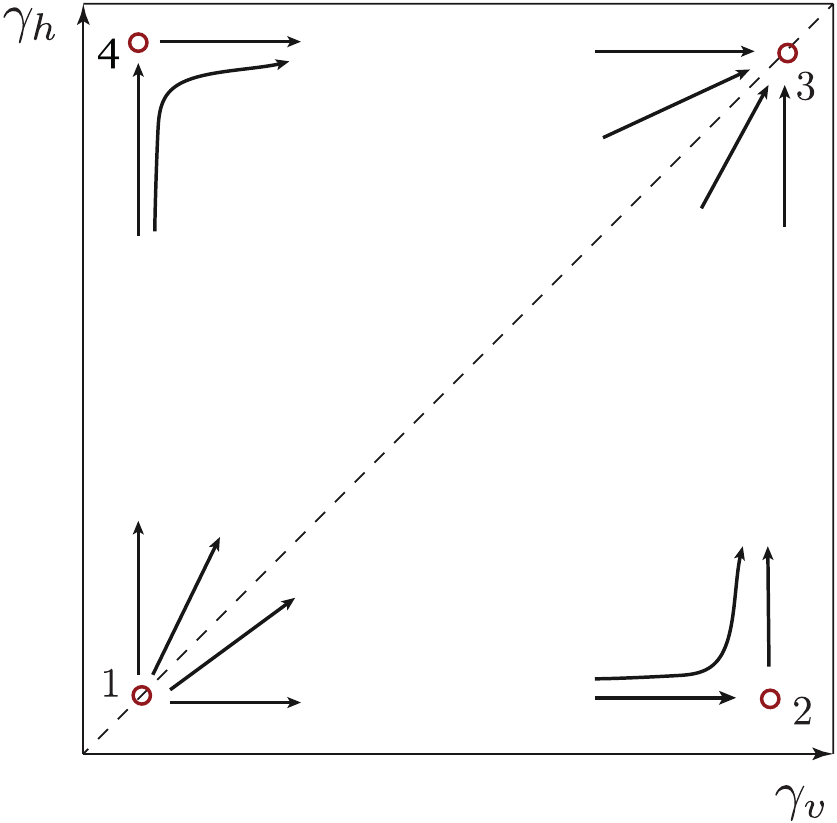}
\caption{Description of the  RG flow in terms of masses $\gamma_h$ and $\gamma_v$, which denote the strength of 
elating tunnelling in horizontal and vertical direction, respectively. Near the RG-unstable weak-weak fixed point 1, there are perturbative corrections to the Gaussian (1/2,1/2) conductance in both $\gamma_h$ and $\gamma_v$, which grow under the RG. In the vicinity of the  weak-strong fixed point 2, instantons in the strong scattering term $\gamma_v$ are irrelevant, constituting a stable direction in parameter space, while the weak scattering amplitude $\gamma_h$ grows under the RG and thus constitutes an unstable direction. The conductance is close to unity in the vertical direction, and perturbatively small in the horizontal direction. In the vicinity of the strong-strong fixed point 3, both vertical and horizontal tunnelings are strong, with RG irrelevant instanton corrections. The conductance is (1/2,1/2), and the fixed point is stable under RG scaling. }
\label{rgflow.fig}
\end{figure}
%
  
\subsection{Effective Geometry} 
As  we will consider qp scattering
between the wires,  we  need to find  a faithful representation
of quasiparticle operators in terms of bosonic constituents. 
For the purpose of establishing unique commutation relations between qp operators acting on different chirals, we will put the four chiral
edges defining our problem onto a single contour, as shown in
Fig.~\ref{fig:ChiralContour}. We now follow the recipe for representing
quasiparticle operators as discussed by Oshikawa, Chamon, Afleck (OCA in the following) in Appendix ~E5 of  Ref.~[\onlinecite{Chamon}]. 
The starting point is a
representation of chiral bosonic field, generalized for the presence
of zero modes. In operator notation,
\begin{align}
  \label{eq:27}
  \hat\varphi_i(x_i)={1\over \sqrt 2}\left(\hat q_i-{2\pi \over 2L} \hat p_i x_i s_i\right) +\phi_i(x_i).
\end{align}
Here, $x_i$ is confined to the $i$th segment of the contour, and $L$
is extension of that contour. The sign factor is
$s_i=1$ for $i=1,2$ and $s_i=-1$ for $i=3,4$. The Fourier
components of the oscillator modes are such that these modes are
algebraically independent on the different contours, and hence they
commute. The role of zero modes and finite momentum modes in
manufacturing a field continuously covering the whole contour is
indicated in Fig.~\ref{fig:ZeroModes}. We note  that the zero modes satisfy
the commutator relation
\begin{align}
  \label{eq:28}
  [\hat q_j,\hat p_k]=g i \delta_{j,k}.
\end{align}
At this point, a comment on normalization is due: we represent our
chiral fields as $\exp(i\sqrt{2} \varphi)\to
\exp[i(q-{2\pi\over L}p x+\sqrt{2}\phi)]$. This differs from the OCA
convention by the factor $\sqrt{1/g}$. OCA have a convention $\exp[i(
\sqrt{g} q + {2\pi \sqrt{g} \over L} p+\phi)]$, and unit commutation relations
$[q,p]=i$. 

When substituted into a functional integral, $\hat q_i \to q_i(t)$
becomes a time dependent field. In the literature, one often finds the
additional dynamical substitution rule, $\hat p_i x_i \to \hat p_i(x_i
-t)$, where, in order to simplify the notation, we have taken the velocity to be unity.  This we should interpret in terms of the interaction picture
dynamics generated by the zero mode Hamiltonian
%
\begin{align}
  \label{eq:29}
  \hat H_0\equiv {\pi\over 4 g  L} \sum_i \hat p_i^2,
\end{align}
%
and the corresponding zero mode action
\begin{align}
  \label{eq:30}
  S_0[q,p]=\int d\tau\,\left({i\over g}  p_i  d_\tau q_i- H_0\right). 
\end{align}
The corresponding equation of motion
\begin{align*}
  d_t q_i&= {\pi\over L} p_i,\cr
 d_t p_i&=0,
\end{align*}
are solved by $q_i(t)=q_i+p_i {\pi\over L} t$, with constant
$(q_i,p_i)$. Substitution into the field definition Eq.~(\ref{eq:27}) 
leads to the dynamical interpretation
\begin{align*}
  \varphi_i(x_i,t) = {1\over \sqrt{2}}\left(q_i -{\pi\over L}p_i(x_i s_i-t)\right)+\phi_i(x_i,t) 
\end{align*}
with time independent $(q_i,p_i)$. At this stage, the meaning of the
sign factors $s_i$ becomes transparent. They make sure that on
contours $1,2$ causality is towards increasing coordinates ('incoming'), and on
contours $3,4$ towards decreasing coordinates ('outgoing').

\subsection{Generalized Klein factors}

Following OCA, we represent a quasiparticle operator obeying proper
\textit{intra}wire commutation relations as
\begin{align}
  \label{eq:31}
  	  \psi_i = e^{iq_i} e^{-i {2\pi \over L} p_i x_i}
  \,e^{i \phi_i},
\end{align} 
as a product of zero mode operators and the finite momentum vertex operators $\exp(i\phi_i)$.
Quasiparticle
\textit{inter}wire commutation relations are now generated by
multiplication with a factor
\begin{align}
  \label{eq:32}
  \eta_i \equiv e^{i{ \pi\over 2} \sum_j \alpha_{ij} p_j},
\end{align}
where the matrix $\alpha$ is given by
\begin{align}
  \label{eq:33}
  \alpha=
  \left(
    \begin{matrix}
      0&1&1&1\cr
      -1&0&1&-1\cr
      -1&-1&0&-1\cr
-1&1&1&0
    \end{matrix}
  \right).
\end{align}
The matrix structure encodes the sequential ordering of segments on
the contour. For example, the first row indicates $1<2,3,4$. The
second $2>1,4$ and $2<3$. 
We define new operators $\Psi_i = \eta_i \psi_i$, which are  designed such that
\begin{align}
\label{eq:PsiComm}
  \Psi_i \Psi_j = \Psi_j \Psi_i \,e^{ i \pi g \alpha_{ij}}.
\end{align}
The derivation of these commutation relations from the current path integral
formalism is detailed in Appendix~\ref{sec:proof_of_eq_eqpsicomm}.

We aim to understand   how the presence of these 'Klein factors' will
interfere with the scaling of the scattering operators. To achieve this, we
define the latter in such a way that the tunneling points sit at $x_i=0$. In
order to avoid ambiguity, we actually remove the tunneling sites a little bit
from zero as indicated in Fig.~\ref{fig:ChiralContour}. This means that the
$p_i x_i$ terms in the quasiparticle amplitudes drop out. 
We are then led to consider the following bilinears:
\begin{align}
  \label{eq:39}
 \bar \Psi_3 \Psi_1 &= e^{i(q_3-q_1)}e^{i{\pi\over 2}
   (p_1+p_3+2p_2+2p_4)} e^{i(\phi_1-\phi_3)}\, e^{-i {\pi\over 2} g},\cr
\bar \Psi_4 \Psi_2 &= e^{i(q_4-q_2)}e^{i{\pi\over 2} (-p_2-p_4)} e^{i(\phi_2-\phi_4)}\, e^{i {\pi\over 2} g},\cr
\bar \Psi_3 \Psi_2 &= 
  e^{i(q_3-q_2)}e^{i{\pi\over 2} (p_3+p_2)}  e^{i(\phi_2-\phi_3)}\,
  e^{-i{\pi\over 2} g},\cr
\bar \Psi_4 \Psi_1 &=e^{i(q_4-q_1)}e^{i{\pi\over2}(p_1+p_4)} 
e^{i(\phi_1-\phi_4)}\,e^{-i {\pi\over 2} g}.
\end{align}
In deriving these relations, we made use of the commutation relations
\begin{align}
  \label{eq:57}
  e^{is {\pi\over 2} p_i} e^{is' q_i}=e^{is' q_i}e^{is {\pi\over 2}
    p_i} e^{iss' {\pi \over 2} g},
\end{align}
where $s,s'=\pm 1$. These bilinears enter the problem through the scattering action 
\begin{align}
	\label{SScattPsi}
	S_s[\bar\Psi,\Psi]&\equiv S_v[\bar\Psi,\Psi]+S_h[\bar\Psi,\Psi],\\
	&S_v[\bar\Psi,\Psi]=\int d\tau \left(\gamma_{13}\bar\Psi_3\Psi_1+\gamma_{24}\bar\Psi_4\Psi_2+\mathrm{h.c}\right),\nonumber\crcr
	&S_h[\bar\Psi,\Psi]=\int d\tau \left(\gamma_{23}\bar\Psi_3\Psi_2+\gamma_{14}\bar\Psi_4\Psi_1+\mathrm{h.c}\right).\nonumber
\end{align}
Another observable relevant to our discussion is the charge on the
dot. Defining the charge density as $s_i\partial_x \varphi_i =
 \sqrt{2}\pi \rho_i$, we get for the charge 
\begin{align}
  \label{eq:40}
  Q_i = {1\over \sqrt 2 \pi}\int_0^L dx\,s_i\partial_x
  \varphi_i=- {p_i\over 2} -{s_i\over \sqrt 2\pi} \phi_i(0),
\end{align}	
Abbreviating as $\phi_i(0)=\phi_i$, we thus
obtain 
\begin{align}
  \label{eq:41}
  Q=-{1\over 2}(p_1+p_2+p_3+p_4) -{1\over \sqrt{2} \pi} (\phi_1+\phi_2-\phi_3-\phi_4) 
\end{align}
for the charge on the dot. (The mnemonic behind the different sign
configuration is that in $\varphi\sim px+\phi$, the zero mode $p$
comes with a spatial coordinate. That one changes sign when we
effectively count $x\to -x$ on the outgoing wires.). 

We finally note that away from the scattering points $x_i$ the action is
quadratic, implying that the field amplitudes $\phi_i(x_i\not=0)$ can be
integrated out. As a result, we obtain a standard dissipative action
%
\begin{align}
  \label{eq:50}
  S_{\rm diss}[\phi]={T\over \pi g}\sum_{i=1}^4 \sum_n |\omega_n|\, |\phi_{i,n}|^2 \ \ ,
\end{align}
%

\subsection{Effective action}

Eqs.~\eqref{eq:39},~\eqref{eq:41}, and~\eqref{eq:50} describe the basic
constituents of the problem, the scattering operators, the charge operator,
and the dissipative action in terms of the twelve fields $\phi_i, q_i$, and
$p_i$. Throughout, we will assume that the charging energy $E_c=e^2/C$ is the
largest energy scale in the problem. Focusing on smaller scales, we may assume
fluctuations of the charge $Q$ to be strongly suppressed, which removes one
degree of freedom from the problem. Assuming $Q$ to be locked, eight more
linear combinations of the above variables may be integrated out without
further approximation, by a procedure detailed in
Appendix~\ref{sec:DerivationEq52}. To define the resulting effective
action, we define the  new set of variables, 
\begin{align} \label{captialphi.eq} \Phi_0 &=
{1\over 2}(\phi_1+\phi_2+\phi_3+\phi_4), \cr   \Phi_1 &= {1\over
2}(\phi_1-\phi_2+\phi_3-\phi_4),\cr  \Phi_2 &= {1\over
2}(\phi_1-\phi_2-\phi_3+\phi_4),\cr  \Phi_3 &= {1\over
2}(\phi_1+\phi_2-\phi_3-\phi_4).  
\end{align}
$\Phi_3\propto Q$ couples to the Coulomb charging action and is gapped out at energy
scales lower than the charging energy, while $\Phi_0$ is a zero mode with free
action \begin{align}     \label{SPhi0}     &\tilde S_0[\Phi_0]\equiv
{T\over 2\pi g}  \sum_n |\omega_n||\Phi_{0,n}|^2. \end{align} The remaining
two fields $\Phi_{1,2}$ are the nontrivial degrees of freedom of the model and
their effective action is given by
\begin{align}   \label{eq:52}
S[\Phi_1,\Phi_2]&=S_0[\Phi_1,\Phi_2]+S_v[\Phi_2]+S_h[\Phi_1],\cr
&S_0[\Phi_1,\Phi_2]= {T\over 2\pi g}  \sum_n \Phi^T_n    \left(
\begin{matrix}       |\omega_n|&-\omega_n\cr       \omega_n &|\omega_n|
\end{matrix}   \right)\Phi_{-n},\cr   &S_v[\Phi_2]=\gamma_{v}\int d\tau
\cos(\sqrt{2} \Phi_2),\cr &S_h[\Phi_1]=\gamma_{h}\int d\tau  \cos(\sqrt{2}
\Phi_1).   
\end{align} 
Here, $\gamma_{v}=|\gamma_{13} + e^{- i \pi g}\,  \gamma_{24}|$
and $\gamma_h=|\gamma_{23} + \gamma_{14}|$ are effective scattering amplitudes
characterizing the backward and forward strength, respectively, cf.
Eq.~\eqref{eq:Svh}. To understand the structure of this result, first notice
that our problem conserves global charge. This is equivalent to the statement
that it is invariant under a uniform static shift of all variables
$\phi_i\to \phi_i + \theta$. Such shifts leave the fields $\Phi_{1,2,3}$
manifestly invariant, while the zero mode action for $\Phi_0$ does not change
provided $\theta=\mathrm{const.}$

Turning to the action of the remaining two fields $\Phi_{1,2}$ it is
straightforward to verify that in the absence of Klein factors, i.e. for
$c$-number valued scattering coefficients $t_{ij}$, the scattering
Hamiltonian~\eqref{ScatteringH} becomes separable once $Q=\mathrm{const.}$ is
frozen out. The action would then decouple into two independent actions of
fields $\Phi_{1,2}$, describing forward ($\gamma_{13}, \gamma_{24}$) and backscattering
($\gamma_{14}, \gamma_{23}$), resp. These actions are obtained from $S$,
Eq.~\eqref{eq:52}, if one ignores the off-diagonal terms in the kernel of
$S_0$. 

The Klein factors account for the non-commutativity of the
scattering operators. In the reduced representation, after integration over
all auxiliary fields, their heritage is the canonical
contribution to the action (the off-diagonal terms in the matrix kernel) i.e.
a term $\sim \int d\tau\, \Phi_1\partial_\tau \Phi_2$ stating non-commutativity 
of the forward and the backward scattering field.

What are the commutation relations between the fields? 
Usually, the commutator between two conjugate variables $[q,p]= i \alpha$ is related to 
an action ${1 \over \alpha} \int dt  p \dot{q}$. Combining the two off-diagonal elements in the 
action Eq.~(\ref{eq:52}), this would lead to $[\Phi_1,\Phi_2]=i \pi g$. However,
this result is incorrect. As usual in a functional integral based approach,
operator commutation relations are obtained by analyzing the effects of time
ordering in the integral, i.e. by subtracting the expressions $\langle \Phi_1(\pm \epsilon)
\Phi_2\rangle$ from each other. The evaluation of these terms must
take the presence of the diagonal dissipative contributions to the action into
account. We then obtain, with $\epsilon \to 0+$,
%
\begin{align}
&   \langle \Phi_1(\epsilon) \Phi_2(0)\rangle - \langle \Phi_1(-\epsilon) \Phi_2(0)\rangle \nonumber \\[.5cm]
&\quad =  \int_{-\infty}^\infty {d \omega \over 2 \pi} \left( e^{i \omega \epsilon} - e^{- i \omega \epsilon} \right) {\pi g \over 2 \omega} \nonumber \\[.5cm]
&\quad =  {i g \over 2} \int_{-\infty}^\infty  d \omega {\sin \omega \over \omega}  =  i {\pi g \over2} \ \ .
\end{align}
%
This implies the commutation relation
\begin{align}
	[\Phi_1,\Phi_2]=i\frac{\pi g}{2},
\end{align}
which will play an important role throughout.

Before ending this section, let us comment on a point that will play a role in
the interpretation of our results below: the effective action~\eqref{eq:52}
describes the physics of the so-called dissipative Hofstadter model 
introduced by Callan and Freed~\cite{CallanFreed}. Recall that the
Hofstadter model describes the physics of a rectangular lattice subject to a
perpendicular magnetic field. Describing nearest neighbor hopping on the
lattice in terms of two operators $p_{12}$ as $\cos(p_{1,2})$, the action of the system
contains a sum of two $
\cos$-operators, where different coupling constants may account for
anisotropy. The presence of a magnetic field implies the  lack of
commutativity, $[p_x,p_y]\not=0$, i.e. a term $\sim \int d\tau \, p_1
\partial_\tau p_2$ in the action, whose coupling constant is a measure of the
field strength. Finally, dissipation may be introduced by including operators
$\sim \sum_n |
\omega_n| (|p_{1,n}|^2+|p_{2,n}|^2)$. Adding everything up, we arrive at an
action equivalent to~\eqref{eq:52} (an identification $p_i \leftrightarrow
\Phi_i$ understood.) More specifically~\eqref{eq:52} describes the DHM fine
tuned to a configuration where field and dissipation strength balance each
other. What motivated Callan and Freed to generalize the non-dissipative
Hofstadter to the presence of dissipation  was the expectation that the
fractal structure of its single particle spectrum might turn into a fractal
pattern of \emph{phase transitions}. Their analysis indeed confirmed that the
phase plane spanned by field and dissipation strength is covered by a fractal
network of phase transition lines separating phases of localization
($\cos$-coupling constants scaling to zero) from phases of de-localizatoin
(diverging $\cos$-amplitudes.) They arrived at that conclusion within a non-perturbative 
analysis resting on an approximate and an exact self-duality
symmetry of the model. The above-mentioned balancing of field and dissipation
strength in our context means that we are cutting through this pattern along a
line parameterized by the Luttinger interaction strength $g$. Later on, we
will confirm within an instanton approach generalizing that
of~\cite{CallanFreed} that for repulsive interactions, $g\ge 1$, we are
generically in a `delocalized' sector of the phase plane, which is another way
of saying that the intermediate scattering fixed point is approached.

\section{The weak-weak fixed point: perturbative  analysis}
\label{sec:weakweak}
  
In this section, we will analyze how initially small coherent tunneling terms
$S_h$, $S_v$ will grow under the RG, depending on the value of the Luttinger
parameter $g$. To leading order in the couplings $\gamma_h$ and $\gamma_v$,
only one of the fields $\Phi_1$ or $\Phi_2$ enters the analysis, and hence we
only need the correlation functions
%
\begin{align}
  \label{eq:61}
  \langle \Phi_{i,n} \Phi_{i,-n} \rangle = {\pi g \over 2} {1\over |\omega_n|} \ .
\end{align}
%
This information allows us to work out the flow equations for $\gamma_h$,
$\gamma_v$. Decomposing the fields $\Phi_i$ as usual into slow and fast modes
and performing a RG transformation, one sees that the exponential of the field
is renormalized by a factor
%
\begin{align}
  \label{eq:62}
  e^{i \sqrt{2} \Phi_i} &\to b e^{i \sqrt{2} \Phi_i} e^{-  T \sum_{n,f} \langle 
    \Phi_{i,n} \Phi_{i,-n}\rangle} \ \ . 
\end{align}
%
Here, the factor $b$ in front of the exponential is due to the rescaling of
the integration measure $d \tau$ in the action Eq.~\eqref{eq:52}, and the
$\sum_f$ runs  over fast frequencies $\Lambda/b<|\omega_n|<\Lambda$. Using the
correlation function Eq.~\eqref{eq:61}, we find
%
 \begin{align}    
 T \sum_{n,f} \langle  
    \Phi_{i,n} \Phi_{i,-n}\rangle
  &  =  {\pi T g\over 2} \sum_{n,f} {1\over |\omega_n|}\cr &= {g\over 2}
      \int_{\Lambda/b}^\Lambda {d\omega\over \omega}  = {g\over2}  \ln b \ \ .
\end{align}
Taking everything together, we find that  the scattering operators scale up
with dimension $1-g/2$,  the result obtained  earlier by us \cite{AGR12}. The
reason for the deviation from KF scaling with dimension $1 - g$ is that
fluctuations in the incoming and outgoing chiral are partially correlated with
each other. This is due to the fact that for energies below the charging
energy of the dot, every incoming charge fluctuation is split equally between
the two outgoing chirals, such that the amount of independent fluctuations is
effectively halved. Thus, the presence of dissipation makes the quadratic
action unstable with respect to coherent scattering even for the hypothetical
case of chiral edges with an "attractive" interaction $1 < g < 2$.

The analysis  described above relates to the point "1" in close vicinity of
the  Gaussian fixed point with $\gamma_h =0$, $\gamma_v = 0$, denoted as
$(0,0)$ in Fig.~5. Due to the scaling dimension $1 - g/2$ of coherent
tunneling terms, the Gaussian fixed point $(0,0)$ (WWFP)  is unstable in the
physical regime of $g \leq 1$. We want to emphasize that the variables
$\gamma_h$, $\gamma_v$ are not conductances for tunneling in horizontal or
vertical direction, but are instead the tunneling amplitudes, which act as
mass terms for the fields $\Phi_1$ and $\Phi_2$ in the regime of strong
coupling.

\section{ The  weak-strong fixed point (WSFP): phase slip-instanton mechanism }
\label{sec:weakstrong}

We now want to assume that one of the two tunnel couplings is much larger than
the other one, say $\Gamma \equiv \gamma_v \gg
\gamma \equiv \gamma_h$. To heuristically understand what is happening
in this limit, let us temporarily go back to a Hamiltonian description in
which we have two tunneling operators
%
\begin{align*}
  \Gamma \cos(\sqrt{2}  \Phi_2),\qquad \gamma \cos(\sqrt{2}  \Phi_1), 
\end{align*}
%
where the phase operators obey the commutation relations $[\Phi_1, \Phi_2] = i
g \pi/2$. To leading order in $\gamma$, the interaction-picture time evolution
of the exponential $\exp{(-  \Gamma
\cos(\sqrt{2}  \Phi_2))}$ under the action of the perturbing operator $\propto
\gamma$ is governed by the commutators of the type  $[\gamma \cos(\sqrt{2}
\Phi_1), \exp{(-  \Gamma \cos(\sqrt{2}  \Phi_2)) }]$. The non-commutativity of $\Phi_{1}$ and $\Phi_2$ implies that  the exponential of $\Phi_1$
acts as a translation operator for $\Phi_2$
by an amount $g \pi$.  How does the system respond to such a translation? A
naive and incorrect argument suggests that such a phase flip  by $\pi$ (in the
case $ g = 1$) of $\sqrt{2}
\Phi_2$ will cost a large amount of energy due to the large magnitude of
$\Gamma$, hence it should be accompanied by a second,  almost immediate
instanton that will add up $-\pi$ or $+\pi$ to the first one, taking the field
configuration of $\Phi_2$ back to a minimum. This argument suggests that
the lowest terms effectively contributing to the perturbation expansion will
be of order $\gamma^2$. The less naive and correct heuristic argument takes
into  account that   the argument of $\cos(\sqrt{2} \Phi_2)$ contains
finite $k$ modes besides the  zero-modes generating the commutation relations
of $\Phi_2$ relative to $\Phi_1$. For this reason, the original phase change
by $g \pi$ due to the Klein factors can be compensated by a kink in the finite
momentum part of $\Phi_2$. This secondary kink enables the field
$\Phi_2$ to settle back into an energetically favorable configuration without
the help of another tunneling event, but it \emph{does} cost a finite amount
of dissipative action. The more quantitative analysis to be detailed
momentarily shows that the primary effect of the added action contribution is
a change of the scaling dimension of $\gamma$ to   $1 - g$, which should be compared to the
dimension  $1- g/2$ characterizing the weak-weak fixed point discussed in the
previous section.

On a more formal level, we will will apply an instanton analysis in $\Phi_2$.
The complementary operator introduced within the duality transformation is
assumed to be perturbatively weak. The instanton approach assumes that the
relevant field configurations are rare phase slips $\phi_2={1\over \sqrt2}
(2l+1)\pi\to {1\over \sqrt2} (2(l\pm 1)+1)\pi$  between $2\pi$-consecutive
minima of  $\cos \sqrt{2} \Phi_2$. Assuming that these events occur at times
$\tau_i$, the phase profile is best described by its time derivative
\begin{align}
  \label{eq:53}
  \partial_\tau\Phi_2&=\sum_{i=1}^N \sqrt 2\pi s_i
  \delta(\tau-\tau_i)\ ,\cr
  i\omega_n\Phi_{2,n}&=\sum_{i=1}^N \sqrt 2\pi s_i e^{-i\omega_n
      \tau_i} \  , 
\end{align}
where $\delta$ is a broadened $\delta$-function, and $s_i=\pm 1$
denote the direction (the charge) of the phase slip. To facilitate the
evaluation of the instanton functional integral,  we use a 
Hubbard-Stratonovich with a field $\Theta_n$ to partially decouple the quadratic term  
 $|\omega_n|  \Phi_{2,n} \Phi_{2,-n}$ according to 
 %
 \begin{eqnarray}
   & &T\sum_n\left({1\over 2\pi g}
    \Phi_{i,n}\Phi_{i,-n} |\omega_n|+ 
    {1\over \pi g}\Phi_{1,n}\Phi_{2,-n}\omega_n\right)\to\cr
& &\to T \sum_n\left( {1\over \pi g}  \Phi_{1,n}\Phi_{1,-n} |\omega_n|+ {\pi
  g\over
  2} \Theta_n\Theta_{-n}|\omega_n| \right.  \nonumber \\
  & &\left.  + \Theta_n \Phi_{1,-n} |\omega_n| -
  \Theta_n
\Phi_{2,-n}\omega_n\right) \ \ .  \label{eq:55}
\end{eqnarray}
%
Doing the Gaussian integral over $\Theta$, one gets from the second to
the first line. We note that $\Theta$ and $\Phi_2$ are a canonical
pair. Denoting the instanton action by $S_i$, the functional integral now
becomes
%
\begin{eqnarray}
  \label{eq:54}
  \mathcal{Z} &=& \int D\Phi_1D\Theta e^{-S_0[\Theta,\Phi_1]-S_\gamma[\Phi_1]}  \\
  & &\hspace*{-.5cm}  \times \sum_N {1\over N!}
  \sum_{\{s_i\}} \int \prod d\tau_i e^{-NS_i} e^{i\sqrt2 \pi  \sum_{i=1}^N s_i T\sum_n \Theta_n
  e^{i\omega_n\tau_i}} \nonumber \\
   &=& \int D\Phi_1D\Theta e^{-S_0[\Theta,\Phi_1]+e^{-S_i} \int d\tau \cos(\sqrt 2 \pi
  \Theta)-S_\gamma[\Phi_1] }.\nonumber
\end{eqnarray}
%
Here,  $S_\gamma[\Phi_1]=\gamma \int d\tau\,\cos(\sqrt 2 \Phi_1)$, and 
%
\begin{eqnarray}
  S_0[\Theta,\Phi_1]&= &T \sum_n\left( {1\over \pi g}  \Phi_{1,n}\Phi_{1,-n} |\omega_n|
 \label{eq:56} \right. \\ 
 & & \left.  + {\pi
  g\over
  2} \Theta_n\Theta_{-n}|\omega_n|      
   +   \Theta_n \Phi_{1,-n}
|\omega_n|\right)  \nonumber\\
&= &  {T\over 2} \sum_n (\Phi_{1,n},\Theta_n)
\left(
  \begin{matrix}
    {2\over \pi g} &1\cr 1 &\pi g
  \end{matrix}
  \right)|\omega_n|
\left(
  \begin{matrix}
    \Phi_{1,-n}\cr \Theta_{-n}
  \end{matrix}
  \right). \nonumber 
\end{eqnarray}
%
We are now in a position to explore scaling dimensions. Due to the
assumed weakness of both $\exp(-S_i)$ and $\gamma$, we can treat the
two operators independently of each other. The basic
formula we use is that for a field $\phi$ with correlation $\langle
\phi_n \phi_{-n}\rangle = {\kappa\over T |\omega_n|}$, we have
\begin{align*}
  e^{ic \phi} \to e^{ic\phi} b^{1-{\kappa c^2\over 2\pi}}  
\end{align*}
in an RG step. In our particular case, $\langle X_n X_{-n}\rangle
={1\over T|\omega_n|}
M^{-1}$, where $M$ is the matrix controlling the free action:
\begin{align}
  \label{eq:59}
  \langle \Phi_{1,n} \Phi_{1,-n}\rangle &= {1\over T |\omega_n|} \pi
  g\Rightarrow \kappa=\pi g,c=\sqrt{2}\cr
 \langle \Theta_{n} \Theta_{-n}\rangle &= {1\over T |\omega_n|}
 {2\over\pi g}\Rightarrow \kappa={2\over \pi g},c=\sqrt{2}\pi.
\end{align}
We then find that the two coupling constants scale as
\begin{align}
  \label{eq:60}
  e^{-S_i}\to e^{-S_i} b^{1-{2\over g}},\qquad \gamma \to \gamma b^{1-g}.
\end{align}
Thus, in the regime $g < 1$, the weak coupling $\gamma$ is a relevant perturbation, while the exponential of the instant on action scales to zero, indicating that instantons in the strongly coupled scattering channel are irrelevant. The weak-strong FP has an unstable direction corresponding to the growth of the "weak" scattering channel, while the irrelevance of instantons in the "strong" channel indicates that the channel stays at strong coupling. This situation corresponds to point "2" (and similarly to point "4")  in the phase diagram Fig.~6.

\section{ The strong-strong fixed point (SSFP): instanton analysis }
\label{sec:strongstrong}

We now want to discuss the situation when both the vertical and horizontal
tunneling terms are strong, such that none can be discussed perturbatively
anymore. In the ground state of the system both fields $\Phi_1$ and
$\Phi_2$ then occupy one of the minima of the $\cos$-terms, and elementary
excitations above the ground state are jumps from one such minimum to the
neighboring minimum, see Fig.~7. In this way, the field configurations move on
a lattice of ground states, which are connected by instantons (jumps) in the
fields. The scaling dimension for such a jump can be obtained by computing the
action for the instanton due to the quadratic dissipative term in the fields,
as described in the section above. In this way, we establish that the strong-strong 
fixed point (SSFP, denoted by the point 3 in Fig.~5) is indeed stable.
In the following, we will present a more formal derivation of this result.

%
\begin{figure}[t]
\vspace*{.5cm}
\includegraphics[width=0.6\linewidth]{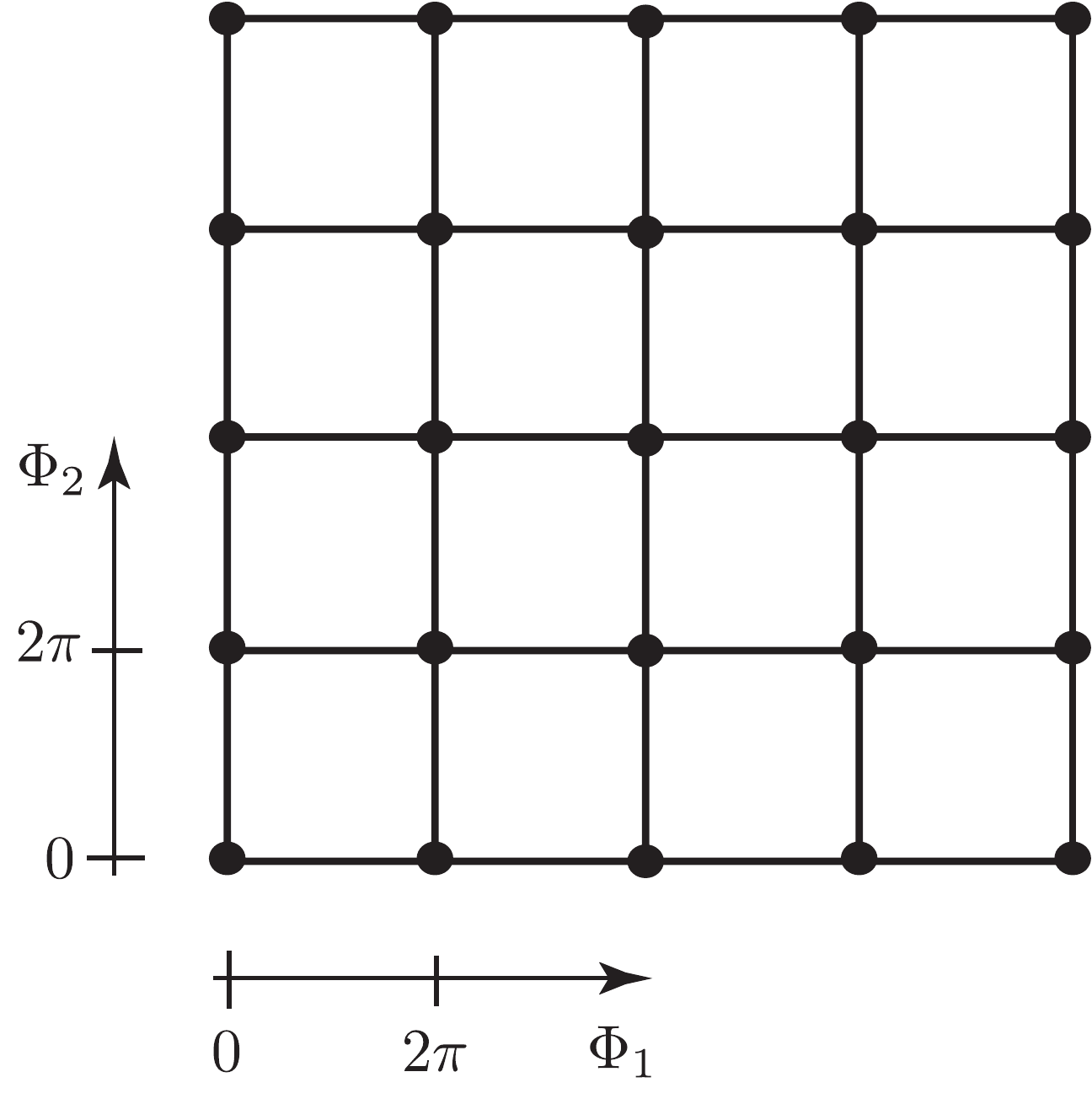}
\caption{Square lattice for instantons in the $(\Phi_1,\Phi_2)$-fields.  Jumps in the horizontal and vertical direction have equal length, and instantons in diagonal direction are less 
relevant than horizontal and vertical ones.   }
\label{fig:6}
\end{figure}
%

If both $\gamma_{v,h}>0$ are strong, we may subject the problem to an
instanton duality mapping. In this regime, the relevant excitations
are rare phase slips $\Phi_i = (2l_i+1)\pi \to (2(l_i\pm 1)+1)\pi$
between consecutive minima of the $\cos$-potentials. Assuming these
events to occur at times $\tau_{i,a}$, the phase profiles
are best described by their time derivative:
\begin{align}
  \label{eq:53a}
  \partial_\tau\Phi_i&=\sum_{a=1}^{N_i} 2\pi s_{i,a}
  \delta(\tau-\tau_{i,a}),\cr
  i\omega_n\Phi_{i,n}&=\sum_{a=1}^{N_i}  2\pi s_{i,a} e^{-i\omega_n
      \tau_{i,a}}.
\end{align}
where $\delta$ is a broadened $\delta$-function, and $s_{i,a}=\pm 1$
denote the direction (the charge) of the phase slip. To facilitate the
evaluation of the instanton functional integral we
Hubbard-Stratonovich decouple the quadratic dependence in $\partial_\tau\Phi_i$
or $i\omega_n \Phi_{i,n}$.
Writing the free Gaussian action as $S_0[\Phi]={T\over 2}
\sum_n \Phi_{n}^T \omega_n X_n^{-1} \Phi_{-n}$, where the matrix $X$ is defined
through Eq.~\eqref{eq:52} as $X_n=\pi g \left(\begin{smallmatrix}
	\mathrm{sgn}(n)&1\crcr -1 &\mathrm{sgn}(n)
\end{smallmatrix}\right)$.
We complete the square in terms of a dual integration variable $\Theta$ as
%
\begin{align}
  \label{eq:63}
  S_0[\Phi]&\to S_0[\Theta,\Phi]\equiv \crcr
 &\equiv {T\over 2}\sum_n\Theta_n \omega_n X_n \Theta_{-n}+T\sum_n\Theta_{n}^T
  \omega_n \Phi_{-n}.
\end{align}
%
The second term tells us that $\Phi$ and $\Theta$ are
canonical variables.

We denote the instanton actions as $S_{\mathrm{i},i}$ and, keeping in mind
their exponential smallness in the coupling constants,
$\gamma_{v,h}$, write down a double expansion 
%
\begin{widetext}
\begin{align}
  \mathcal{Z}&=\int D\Theta \, e^{-{T\over 2}\sum_n\Theta_n \omega_n
    X_n \Theta_{-n}} \sum_{N_i} {1\over N_1! N_2!}
   \sum_{\{s_{i,a}\}}
\int \prod d\tau_{i,a} \,e^{-\sum_i(N_i
       S_{\mathrm{i},i}+i 2\pi \sum_a s_{i,a} T\sum_n e^{i\omega_n
         \tau_{i,a}}\Theta_{i,n})}\nonumber \\[.5cm]
         \label{eq:65}
&=\int D\Theta \,e^{-{T\over 2(2\pi)^2 }\sum_n\Theta_n \omega_n
    X_n \Theta_{-n}-\sum_i  e^{-S_\mathrm{i},i} \int d\tau\,\cos(\Theta_i)}.
\end{align}
\end{widetext}
%
In the last step, we scaled $\Theta$ by a factor $\pi$ for convenience. 
Substituting the form of the $X$-matrix, we arrive at an
action
%
\begin{eqnarray}
  \label{eq:66}
  S[\Theta] &= &{T\over 2}  {g\over 4\pi} \sum_n
  \Theta_{n}^T
  \left(
    \begin{matrix}
      |\omega_n|&\omega_n\cr -\omega_n&|\omega_n|
    \end{matrix}
  \right)\Theta_{-n}\\
  & & +  \sum_i e^{-S_\mathrm{i},i} \int d\tau\, \cos(\Theta_i). \nonumber
\end{eqnarray}
%
dual to \eqref{eq:52}. The
  differences are in the coupling constants: $g \to 
  2/g$, and $\gamma_{v/h}\to e^{-S_\mathrm{i},2/1}$. This means that for
  very large \textit{and nearly equal} coupling constants $\gamma_i$,
  the coupling constants of the dual theory can be renormalized by
  first order RG analysis where the reasoning of the foregoing section
  obtains the scaling
  \begin{align}
    \label{eq:64}
    e^{-S_{\mathrm{i},i}}\to e^{-S_{\mathrm{i},i}}b^{1-{1\over
        g}}.
  \end{align}
At first sight, this result looks surprising: for $g<1$ the instanton
insertions are irrelevant, and the system flows to strong
coupling, while for $1<g<2$ the instanton operators are relevant,
and the weak potential perturbation addressed in the previous
section is also relevant. This suggest the existence of an
intermediate fixed point. Finally, for $g=1$, the instanton operators
are marginal, a situation which seems inconclusive at first. 

\begin{figure}[t]
\vspace*{.5cm}
\includegraphics[width=0.8\linewidth]{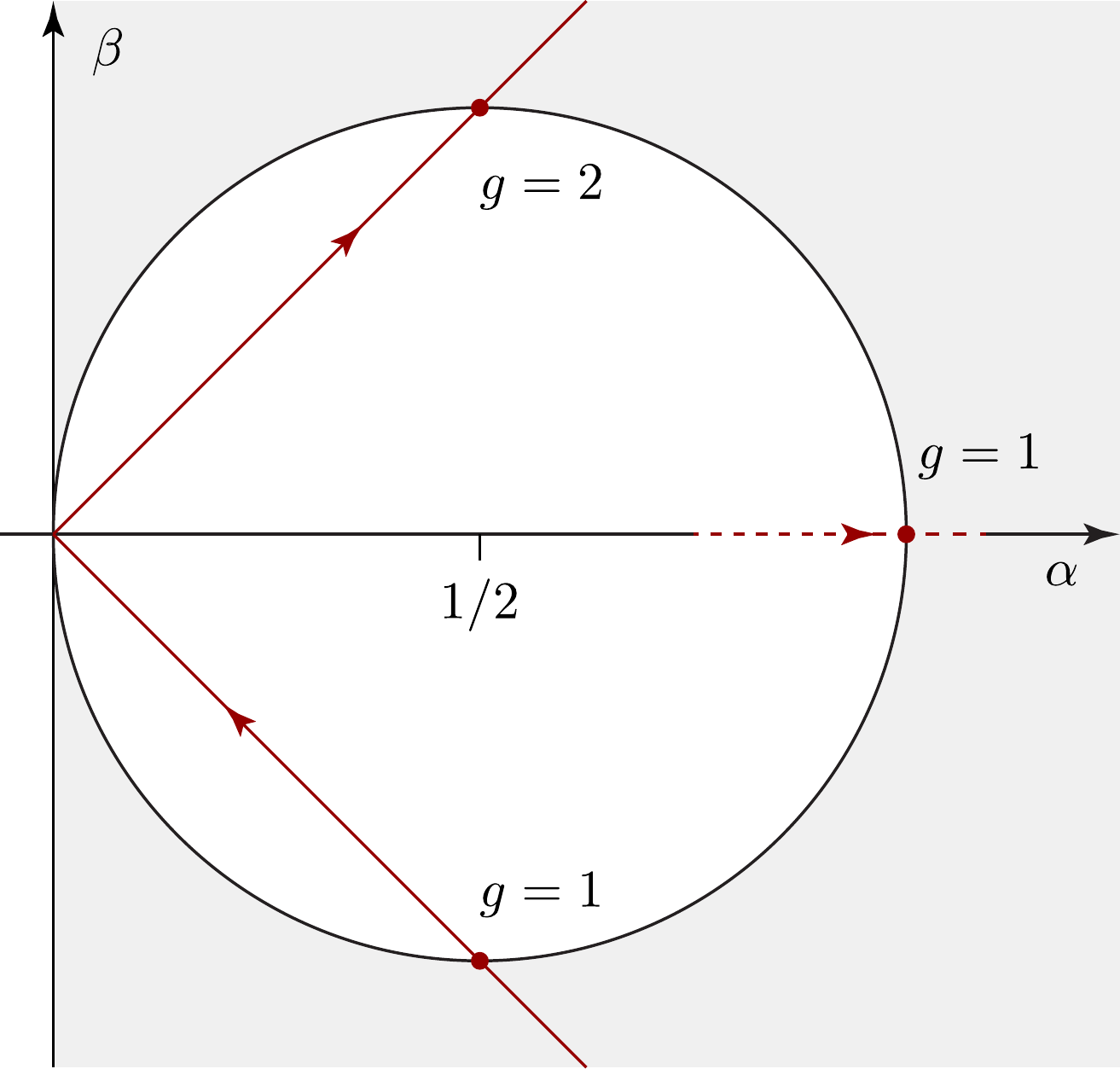}
\caption{Relevancy of perturbations as a function of the two parameters $\alpha,
\beta$ defined in the text. The white disk/shaded area is the region of the phase diagram where scattering terms are irrelevant/relevant. At the weak coupling fixed point, $\alpha=\beta=g^{-
1}$ and the region of relevant perturbation (shaded) is reached as
$g\searrow 2$ (line departing from origin corresponding to lowering $g$). At
strong coupling $\alpha =-\beta=g/2$ and the region of irrelevant perturbation
(circle) is reached for $ g\searrow 1$ (on the line $\beta = - \alpha$.) The line
$\beta=0$ corresponds to the KF model, where perturbations around weak
coupling become relevant at $g=1$. This means that for repulsive lead
interactions, $g<1$, perturbations at weak (strong) coupling are relevant
(irrelevant) implying the stability of the stability of the strong coupling
fixed point.}
\label{fig:6a}
\end{figure}

In order to clarify this situation in more depth,  we follow Callan \& Freed
\cite{CallanFreed}
and generalize the matrix $X_n$ as 
\begin{align}
	X_n^{-1}\to \frac{1}{2\pi}\left(\begin{matrix}
		\alpha\,\mathrm{sgn}(n)&\beta \crcr -\beta &\alpha\,\mathrm{sgn}(n)
	\end{matrix} \right),
\end{align}
i.e.~we allow
for independent coefficients of dissipation ($\alpha$) and canonical
term ($\beta$). Our current
situation is recovered by setting $\alpha=\beta = {1\over g}$. With
this generalization,  we have $\langle\Phi_{i,n} \Phi_{i,-n}\rangle =
{2\pi\over T} {1\over |\omega_n|}{\alpha\over \alpha^2 + \beta^2}$,
and the perturbative scaling dimension becomes $1-{\alpha\over
  \alpha^2 + \beta^2}$. This tells us that the weak scattering
operators are relevant if $\alpha < \alpha^2+\beta^2$. Defining
$z=\alpha+i\beta$, we realize that $z$ must lie outside a circle in
the complex plane centered around $(1/2,0)$ and with radius
$1/2$ (cf. Fig.~\ref{fig:6a}.) In our case, we are sitting on a line $z={1\over g}(1+i)$, and
for $g<2$ we are outside that circle, i.e.~ we are dealing with relevant
operators. We note that for $\beta=1$, the condition collapses to
$g<1$, corresponding to  the standard KF case. 

In the dual case, we have the same situation, only that $X_n^{-1}$ gets
replaced by $X_n^{-1}\to {2\pi\over \alpha^2 +
  \beta^2} \left(\begin{smallmatrix}
  	\alpha\,\mathrm{sgn}(n)&-\beta \crcr \beta &\alpha\,\mathrm{sgn}(n)
  \end{smallmatrix} \right)$.  Defining
$\tilde z = {\alpha\over \alpha^2+\beta^2} -i {\beta\over
  \alpha^2+\beta^2}$, we realize that the instanton duality amounts to
a mapping $z\to \tilde z = z^{-1}$. The instanton operators are irrelevant,
provided $\tilde z$ lies \textit{inside the circle} above. In our present
setting, this requires that $1/(g^{-1}(1+i)={g\over 2}(1-i)$ lie inside the
circle. For $g<1$ this is the case. 
In the hypothetical case of attractive lead interactions, $g>1$ we are outside and the instanton operators are
relevant, while the weak scattering operators are also relevant. 

For $g=1$, we are borderline. We note that backscattering is a relevant
 perturbation in the weak-weak limit - with a scaling dimension of the
 backscattering probability equal one.  This is in line with
 reference~[\onlinecite{ABG}] (cf. Eq.~(346) with $N_{ch}=2$, see also
 reference [\onlinecite{Matveev}]). In the strong-weak and in the
 strong-strong limit the flow of the backscattering probability  is marginal.
 This is indeed  verified in the next section (cf. also Fig.~10). For this
 behavior to hold, we have utilized in our analysis the instantaneous
 character of backscattering. Reference~[\onlinecite{ABG}] allows for more
 general scenarios which may lead to truncation of the scaling flows
 (cf.~Eq.~(348) there).

%
%

In  the next section, we treat the model for the specific value 
$g=1$  of the 
Luttinger parameter by using the method of  fermionization, and find agreement with the scenario discussed here. 
Even if we  consider non-interacting leads, yet,
there is still interaction in the model: the electrostatic charging energy,
which couples the four chirals among each other.  It is clear from the scaling
obtained above that the flows in the vicinity of both the WSFP/SWFP and  the
SSFP are marginal. This nicely reproduces the expectation that for Fermi-
liquid leads, scattering should be marginal in the low-energy limit, and that
arbitrary ratios between forward and backward scattering strength can be
realized.

\section{General comments on the  analysis}

We would like to discuss some subtleties of the analysis presented above:  (i)
We assume a cutoff (voltage or temperature), which may be  small, yet larger
than the level spacing $\Delta$ of the QD (which for a typical quantum Hall
geometry may be  exceedingly small.) Had we pursued the RG all the way down to $\Delta$,  the character of
the model would have changed: there are no inelastic excitations then, hence
the inelastic channel is frozen. In that case, in the presence of only elastic
scattering, we are driven to either full transmission or full reflection, in
agreement with the Kane-Fisher result. In the present analysis, we
formally take the limit of an infinite QD, before  pursuing our RG procedure
all the way down to zero temperature or zero voltage. (ii)    
In the absence of forward scattering channels ($\gamma_{14}= \gamma_{23} = 0$), our model and the results obtained 
for its scaling reduce to those  
considered by   Furusaki-Matveev \cite{}. (iii)   
Except for the highly degenerate points of perfect scattering amplitude cancellation 
$\gamma_{13} + e^{- i \pi g} \gamma_{24}=0$ or $\gamma_{23} + \gamma_{14}=0$, our analysis  holds regardless of scattering phase shifts and strengths
(at the degenerate point, the model collapses to a variant of the Kane-Fisher problem). 
However, these points require fine tuning and are nongeneric.
The specific conditions for perfect scattering amplitude cancellations are  related to the choice of the Aharonov-Bohm phase enclosed by the the
corresponding  tunneling path. Here we have followed  special choice, that of
a zero  Aharonov-Bohm flux. We   note that OCA    Ref.~\onlinecite{Chamon} (in their 3 lead geometry) have
allowed  the freedom of choosing the Aharonov-Bohm flux. (iv)    An extension
of our   analysis from FQHE to interacting Luttinger wires is presented in
Appendix

\section{ Refermionization for the case $g=1$ }

\begin{figure}[h]
  \centering
  \includegraphics[width=7cm]{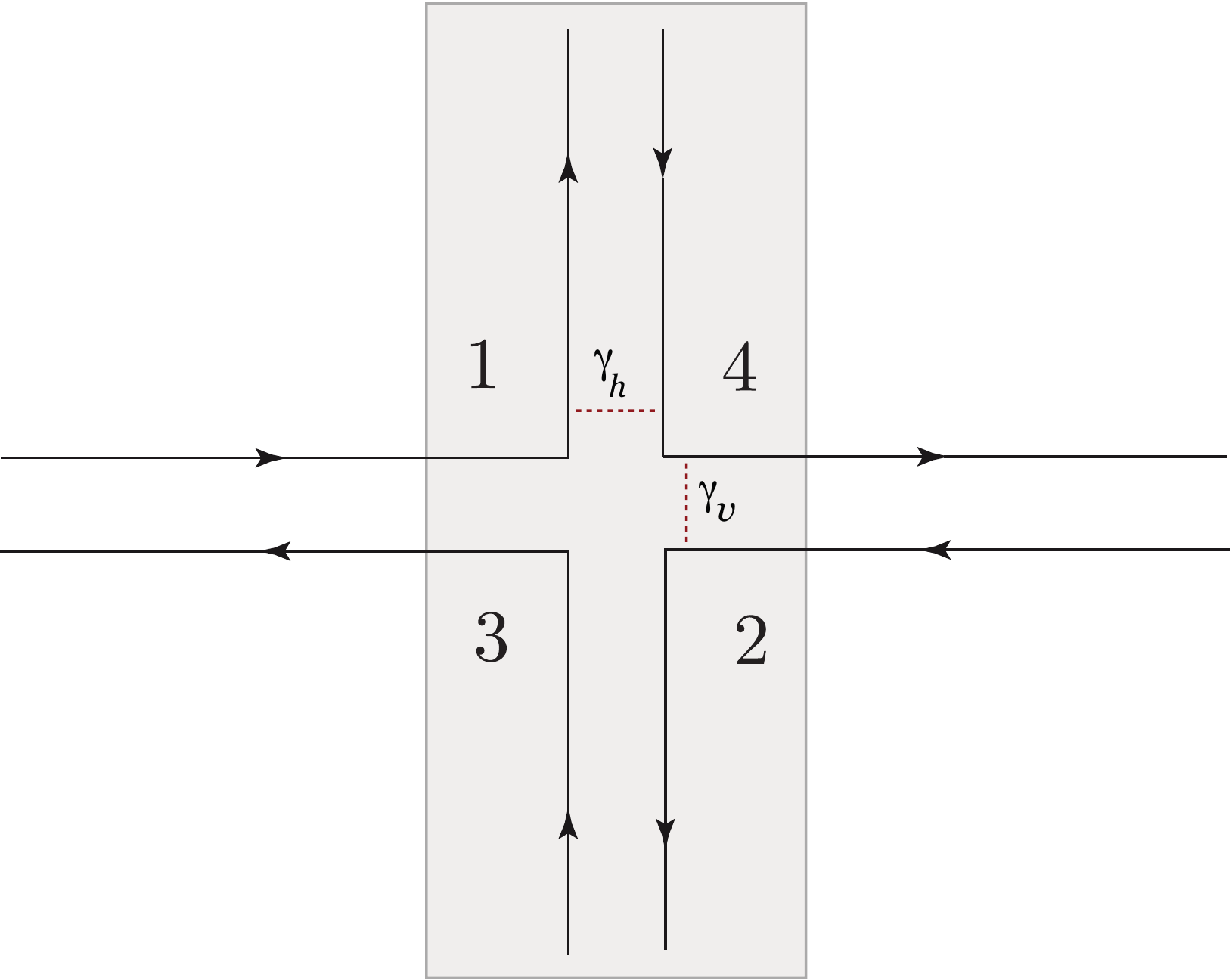}
  \caption{Setup with two competing scattering amplitudes 
  $\gamma_{14} \equiv \gamma_h$ and $\gamma_{24} \equiv \gamma_v$. 
  We consider the special case of non-interacting leads $g=1$, and denote the 
  probability for elastic scattering from segment $1$ to segment $4$ by $P_{14}$, and the probability 
  for elastic scattering from segment $2$ to segment $4$ by $P_{24}$. 
  }
  \label{fig:refermionization_setup}
\end{figure}

We consider a special model,  where there are only two nonzero scattering amplitudes $\gamma_{14}\equiv \gamma_h$ and 
$\gamma_{24}\equiv \gamma_v$, see Fig.~\ref{fig:refermionization_setup}, and where the leads are non-interacting. The model has an intereting dynamics all the same, since at low energies the mode $\Phi_3$ describing charge fluctuations in the quantum dot is already frozen. In this limit, the tunneling amplitudes $\gamma_h$ and $\gamma_v$ are both relevant perturbations. Using a mapping onto non-interacting fermions, we derive an exact solution of the model,  which describes the competition between
the competing scattering processes $\gamma_h$ and $\gamma_v$.  We find that at high temperatures and small $\gamma_h$ and $\gamma_v$, both $\gamma_h$ and $\gamma_v$ increase $\propto T^{-1/2}$, in agreement with a weak coupling perturbative RG in Section IV. The asymptotic value for $T\to0$ is non-universal and depends on the ratio of initial values of $\gamma_h$ and $\gamma_v$. This finding 
complements the results of Sections \ref{sec:weakstrong} and \ref{sec:strongstrong}, where it was found that non-interacting leads with $g=1$ separate two different regimes of RG flow.

%
%

The starting point for our discussion are the tunnelling terms and the dissipative term  in   Eq.~(\ref{eq:52}). Due to the off-diagonal terms in $S_0$, the fields $\Phi_1$ and $\Phi_2$ satisfy the commutation relation $[\Phi_1,\Phi_2]= i \pi {g\over2}$. In this appendix, we specialise to non-interacting leads with $g=1$. Due to the presence of the charging energy on the dot, the diagonal entries of $S_0$ describing dissipative processes are given by $|\omega_n|/2 \pi$. We compare this value with the dissipative action $|\omega_n|/4 \pi \nu$  for the charge density field describing backscattering of quasi-particles (in the absence of a QD) between two counter-propagating fractional quantum Hall edges with Luttinger parameter $\nu$. 
We find that non-interacting leads with a  QD charging energy have a dissipative action equivalent to that of interacting $\nu=1/2$ 
LL edges in the absence of a QD. The special case of backscattering between $\nu=1/2$ LL edge states is can be solved exactly by the method of refermionization \cite{Chamon+96}, which we will adapt to our problem of two competing scattering processes in the following.  
In order to achieve this goal, we undo the shift Eq.~(\ref{eq:44}), such that in the following the fields $\Phi_1$ and $\Phi_2$ contain finite frequency modes only. The non-commutativity between the fields is now described by Majorana operators $\eta_1$, $\eta_3$, and $\eta_4$ with $\{\eta_i, \eta_j\} = 2 \delta_{ij}$. Switching to a Hamiltonian formalism suitable for refermionization, the tunnelling Hamiltonian is given by 
%
\begin{eqnarray}
H_{\rm tun} & = &   i \eta_1 \eta_4 \gamma_h  \left[  e^{i \sqrt{2}  \Phi_1(0)} \ + \  e^{- i \sqrt{2} \Phi_1(0)}  \right]  \\
& &  + 
i \eta_2 \eta_4 \gamma_v  \left[ e^{i \sqrt{2} \Phi_2(0)} \ + \  e^{- i \sqrt{2} \Phi_2(0)}  \right]
  \ \ . \nonumber
\end{eqnarray}
%
Here, we have absorbed a factor $1/2$ into the tunnelling amplitudes $\gamma_h$ and $\gamma_v$ as compared to Eq.~(\ref{eq:52}). Taking into account that the dissipative actions of $\Phi_1,\Phi_2$ are equal to that of an infinite {\em chiral} Luttinger liquid, the operators
$e^{i \sqrt{2} \Phi_{1/2}} $ can be interpreted as the operator $\Psi_{1/2}$ of  free chiral electrons. However, to make this equivalence more precise, we 
need to introduce  Majorana fermions $f_{1/2}$ as a Klein factor for the new electrons \cite{Chamon+96}, and define
%
\begin{equation} 
e^{i \sqrt{2} \Phi_{1/2}(0)} = \sqrt{2 \pi a} \  \Psi_{1/2}(0) \ f_{1/2}  \ \ .
\end{equation}
%
Here, $a$ denotes the short distance cutoff of the theory, which is needed to make sure that the fermion fields $\Psi_{1/2}$ have the proper dimension of one over square root of length. 
In this way, we obtain the fermionized tunnel Hamiltonian 
%
\begin{eqnarray}
H_{\rm tun} & = & i \eta_2 \eta_4 \gamma_v \sqrt{2 \pi a}  \left[  \Psi_2(0)\  f_2 \ + \   f_2 \  \Psi_2^\dagger(0) \right] \\
& &  +   i \eta_1 \eta_4 \gamma_h \sqrt{2 \pi a}  \left[  \Psi_1(0) \  f_1 \ +  \  f_1 \ \Psi_1^\dagger(0) \right] 
\ \ . \nonumber
\end{eqnarray}
%
In addition,  the fermions have a kinetic term (the velocity is taken to be unity) 
%
\begin{equation}
H_0 \ = \ \int dx \left[   \Psi_1^\dagger(x)  \left( - i \partial_x \right) \Psi_1(x)   \ + \  \Psi_2^\dagger(x)  \left( - i \partial_x \right) \Psi_2(x)\right]   . 
\end{equation}
%

Since the leads are non-interacting, we can discuss transport through the QD in terms of a scattering formalism. For an incoming particle on 
edge $1$, we define the probability $P_{14}$ to scatter onto the outgoing edge mode $4$. Similarly, for an incoming particle on edge $2$, we denote the probability to scatter onto the outgoing edge mode $4$ by $P_{24}$.  Then, in Appendix D we derive the result 
%
\begin{figure}[h]
\includegraphics[width=0.8\linewidth]{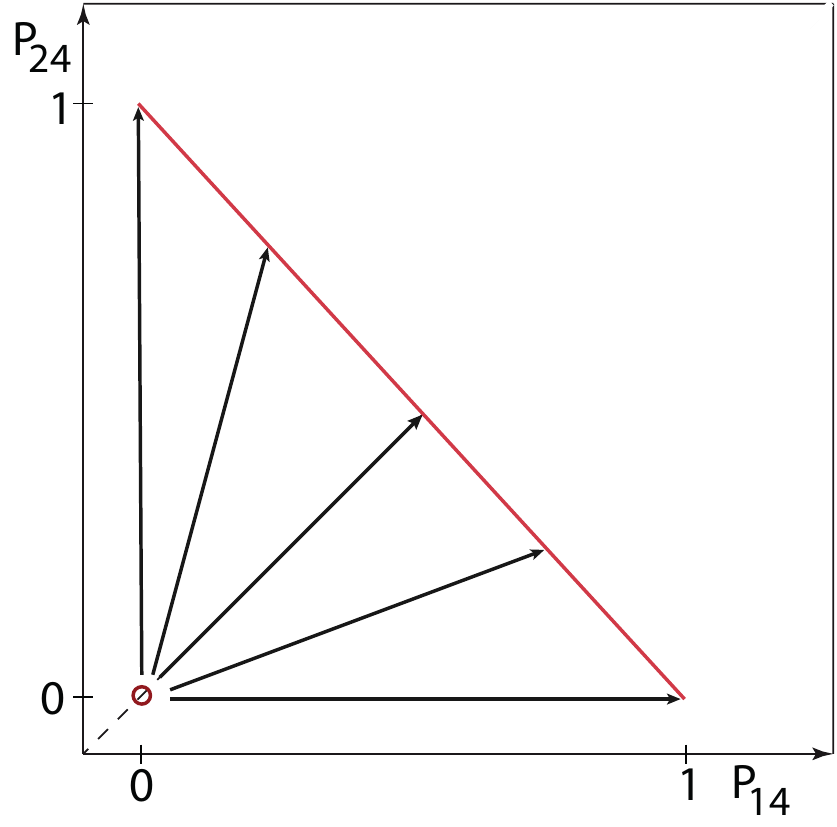}
\caption{Description of the  RG flow in terms of  scattering probabilities $P_{14}$ and $P_{24}$, which in the weak coupling limit are 
proportional to the square of tunneling amplitudes $\gamma_h^2$ and
$\gamma_v^2$, respectively.  Near the RG-unstable weak-weak fixed point (red
circle), there are perturbative corrections to the Gaussian (1/2,1/2)
conductance in both $\gamma_h$ and $\gamma_v$, which grow under the RG. The RG
flow terminates on a line of fixed points (red line), which are characterized
by scattering probabilities and a conductance which depend on the ratio
$\gamma_v/\gamma_h$ of bare scattering amplitudes. }
\label{rgflow_fermionic.fig}
\end{figure}
%
%
\begin{subequations}
\begin{eqnarray}
P_{14} & = & {\Gamma_h \over \Gamma} \ F\left({T \over \Gamma}\right) \label{P_14.eq}\\[.5cm]
P_{24} & = & {\Gamma_v \over \Gamma} \ F\left({T \over \Gamma}\right) \ \ , \label{P_24.eq}
\end{eqnarray}
\end{subequations}
%
with 
%
\begin{equation}
\Gamma_h \ = \ 4 \pi a \gamma_h^2 \ \ , \ \ \ \ \Gamma_v \ = \ 4 \pi a \gamma_v^2 \ \ , \ \ \ \  \Gamma \ = \ \Gamma_h \ + \ \Gamma_v \ \ , 
\end{equation}
%
and 
%
\begin{equation}
 F(x) = {2 \over x} \arctan{x\over 2} \ \ .
 \end{equation}
 %
The scaling function has the limiting behaviors $F(x) = 1 - {1 \over 3} (x / 2)^2$ for small $x$, and $ F(x) = {2 \over x}$ 
for large $x$. Using these asymptotics, we find 
%
\begin{eqnarray}
P_{14} & = & \left\{ 
\begin{array}{cc}   {2 \Gamma_h \over T} & {\rm for } \ T \gg  \Gamma \\[.5cm]
{\Gamma_h \over \Gamma}  & {\rm for} \ {T \over \Gamma} \to 0 
\end{array} \right. \ \ , 
\end{eqnarray}
%
and similarly for $P_{24}$, with $\Gamma_h$ and $\Gamma_v$ interchanged. In
the high temperature limit, this result agrees with the perturbative analysis
presented in Section \ref{sec:weakweak}. Since $P_{14} \propto \gamma_h^2$ in
the high temperature limit,  the flow of $P_{14}$ is representativ  for the
flow of the tunneling amplitude $\gamma_h$, and similarly for $P_{24}$ and
$\gamma_v$. The flow of $P_{14}$ and $P_{24}$ is shown in
Fig.~\ref{rgflow_fermionic.fig}. For weak bare values of $\gamma_h$,
$\gamma_v$ with $\Gamma_h/T \ll 1$, $\Gamma_v /T \ll 1$, the flow starts in
the vicinity of the unstable weak-weak fixed point. In terms of the variables
$P_{14}$ and $P_{24}$, the flow occurs along straight line trajectories, which
stop at the line of fixed points defined by $P_{14} + P_{24} = 1$. The end
point of the flow on this fixed line is determined by the ratio of initial
couplings $\gamma_h/\gamma_v$. In particular, in the weak-strong limit with,
say, $\gamma_v \gg \gamma_h$, the flow is truncated at a temperature $T
\approx \Gamma_v$, and the weak coupling $\gamma_h$ is marginal. This
situation is intermediate between the case $g > 1$, where the weak coupling is
irrelevant, and the case $g < 1$, where the weak coupling is relevant. Thus,
the exact solution for the case of noninteracting leads with $g=1$ is in full
agreement with the perturbative analysis presented in Section
\ref{sec:weakstrong}. We find that for $g=1$ there is no strong-strong fixed
point, which again is in agreement with the analysis in Section
\ref{sec:strongstrong}, in which it was found that $g=1$ separates the case of
a stable strong-strong fixed point for repulsive lead interactions with $g <
1$ from the case of an unstable strong-strong fixed point for $g > 1$.

\section{  Summary }

In summary, we have considered a model of scattering between one-dimensional
chiral leads in the presence of both inelastic and elastic scattering
channels. We have found that this combination of scattering can stabilize a
new $1\over 2$ - $1 \over 2$ fixed point, with probability $1/2$ for the
transmission and probability $1/2$ for the reflection of an incoming particle.
This $1 \over 2$ - $1\over 2$ fixed point is intermediate between the 1 - 0
and 0 - 1 fixed points in the presence of only elastic scattering channels. In
order to establish the existence of this fixed point, we employed a non-
perturbative instanton analysis in either one or both of the scattering
channels. Our main result is that the intermediate fixed point is stable for
Luttinger parameters $g <1$ in the leads. This conclusion is backed up by a
refermionization analysis for the special value $g=1$. For non-interacting
fermions with $g=1$, we  recover the well-known marginal relevance of
scattering. While this   $1/2-1/2$ fixed point is the main result of the
present analysis, we recall  that our previous paper \cite{AGR12}  pointed
out novel results concerning noise and current-currnet correlations.

\acknowledgements

We acknowledge useful discussions with P. Brouwer, B. Halperin,  M. Heiblum, and C. Marcus. 
This work was supported by GIF,
 BSF,  SFB/TR 12 of the Deutsche Forschungsgemeinschaft, and DFG grants RO 2247/7-1 and RO 2247/8-1. 

\appendix

\numberwithin{equation}{section}

\section{Proof of Eq.~\eqref{eq:PsiComm}}
\label{sec:proof_of_eq_eqpsicomm}

As usual, the path integral fixes operator ordering through
time order. Consider, for example, two quasiparticle operators, 
%
\begin{eqnarray}
  \Psi_i(\tau+\delta \tau) \Psi_j(\tau) &\to & e^{i\left({\pi\over 2} \sum_{i'}
    \alpha_{ii'} p_{i'}(\tau+\delta \tau) -  q_i(\tau+\delta \tau)\right)  } \nonumber \\
    & & \times e^{i \left(
    {\pi\over 2} \sum_{j'}
    \alpha_{jj'} p_{j'}(\tau) -  q_j(\tau)\right)} \nonumber \\
    & & \times \tilde{\psi_i}(\tau + \delta \tau) \tilde{\psi_i}(\tau) \ \ .
\end{eqnarray}
%
The expression we want to compare with looks the same, except for an
exchange of time arguments $(\tau-\delta \tau)\leftrightarrow
\tau$. Now, imagine this expression inserted in the functional
integral. We may always integrate over $\{q_i\}$,  as these fields
enter the action linearly (assuming a perturbative approach, where all
$\Psi_k$ are expanded out of the exponent.) Integration over the $q$'s
generates the step function profiles
\begin{align*}
  p_k(\tau) = p_{k,0}+g \sum_a s_a \Theta(\tau-\tau_a),
\end{align*}
where $p_{k,0}$ is the initial value, the sum runs over all
appearances of $q_k$ in the exponents, $\tau_a$ are the respective
times, and $s_a=1$ for a $-iq$ and $-1$ for a $+iq$ factor. Now, with
these structures in place, we can explore the behavior of the exponent
above. Denoting by $\tilde{\Phi}_{ij} $ the cumulative phase which will not respond
to an exchange of the time arguments, we have
\begin{align*}
   \Psi_i(\tau+\delta \tau) \Psi_j(\tau)&\to e^{i\tilde{\Phi}_{ij}+ i{\pi g\over
       2}\alpha_{ij}},\cr
 \Psi_i(\tau) \Psi_j(\tau+\delta \tau)&\to e^{i\tilde{\Phi}_{ij} +i{\pi g\over 2} \alpha_{ji}},
\end{align*}
where the phase in the first line reflects the fact that at time
$\tau$ $p_j$ jumped by $g/\sqrt{2}$ and this phase change is read out by the
contribution $\alpha_{ij} p_j(\tau+\delta\tau)$ to the first
phase. Similarly for the second line. The relative phase is given by
\begin{align*}
  e^{i {\pi g\over 2} (\alpha_{ij}-\alpha_{ji})} =  e^{i \pi g \alpha_{ij}},
\end{align*}
as required.

\section{Derivation of the effective action~\eqref{eq:52}}
\label{sec:DerivationEq52}

In this section, we derive the effective three-variable action~\eqref{eq:52}
from the original description in terms of twelve fields, $\phi_i,q_i,p_i$. We start by casting the orthogonal transformation~\eqref{captialphi.eq} into the  more compact notation
%
\begin{align*}
  \Phi=M \phi,\qquad M={1\over 2}
  \left(
    \begin{matrix}
      1&1&1&1\cr	
      1&-1&1&-1\cr
      1&-1&-1&1\cr
      1&1&-1&-1
    \end{matrix}
  \right).
\end{align*}
%
This transformation is orthogonal, $M^T=M^{-1}$, and thus the inverse transformation is given by 
$\phi=M^{-1} \Phi = M^T \Phi$. For the zero modes, this allows to 
define $p=M^{-1}P$ and $q=M^{-1}Q$ as above to obtain
%
\begin{align}
  \label{eq:42}
  Q= -2 P_0 -{\sqrt{2}\over \pi} \Phi_3.
\end{align}
%
Then, the zero mode part of the scattering operators assumes the form
%
\begin{align}
  \label{eq:43}
   \bar \Psi_3 \Psi_1 &= e^{i(-Q_2-Q_3)}e^{i{\pi\over 2} (-P_1+3 P_0)} \bar \psi_3 \psi_1 e^{-i {\pi\over 2} g},\cr
\bar \Psi_4 \Psi_2 &= e^{i(+Q_2-Q_3)}e^{i{\pi\over 2} (+P_1-P_0)} \bar \psi_4 \psi_2 e^{i {\pi\over 2} g},\cr
\bar \Psi_3 \Psi_2 &= 
  e^{i(+Q_1-Q_3)}e^{i{\pi\over 2} (-P_2+P_0)}  \bar \psi_3 \psi_2 e^{-i {\pi\over 2} g},\cr
\bar \Psi_4 \Psi_1 &=e^{i(-Q_1-Q_3)}e^{i{\pi\over2}(+P_2+P_0)} \bar \psi_4 \psi_1 e^{-i {\pi\over 2} g}.
\end{align}
%
The transformations $(q,p)\to (MQ,MP)$ is orthogonal which means that
the canonical piece of the action remains unaltered, as
\begin{align}
  \label{eq:37}
  {i\over g}\int d\tau P_i d_\tau Q_i.
\end{align}
In principle, there is also the zero mode charging term $H\sim \sum_i
P_i^2$ to consider, however, it does not play any role due to the prefactor of inverse contour  length, which 
becomes zero in the thermodynamic limit.

We note that the scattering terms do not couple to
$Q_0$. This means that we may do the integral over $Q_0$ to conclude
that $d_t P_0=\mathrm{const.}$: the total charge on the system is
constant. Without loss of generality, we may call this constant
$0$. Our scattering terms thus simplify to 
\begin{align}
  \label{eq:43}
   \bar \Psi_3 \Psi_1 &= e^{i(-Q_2-Q_3)}e^{-i{\pi\over 2} P_1} \bar \psi_3 \psi_1 e^{-i {\pi\over 2} g},\cr
\bar \Psi_4 \Psi_2 &= e^{i(+Q_2-Q_3)}e^{+i{\pi\over 2} P_1} \bar \psi_4 \psi_2 e^{i {\pi\over 2} g},\cr
\bar \Psi_3 \Psi_2 &= 
  e^{i(+Q_1-Q_3)}e^{-i{\pi\over 2} P_2}  \bar \psi_3 \psi_2 e^{-i {\pi\over 2} g},\cr
\bar \Psi_4 \Psi_1 &=e^{i(-Q_1-Q_3)}e^{+i{\pi\over2}P_2} \bar \psi_4 \psi_1 e^{-i {\pi\over 2} g}.
\end{align}
Next, ignoring the overall charging term, we observe that $P_3$ does
not appear in the scattering operators, nor in the dot charging
term. This leads to the constraint $d_t Q_3=0$, and we may set $Q_3=0$
without loss of generality. As a result we have the further
simplification, down to  
%
\begin{subequations}
\begin{align}
  \label{eq:44a}
   \bar \Psi_3 \Psi_1 &= e^{-iQ_2}e^{-i{\pi\over 2} P_1} \bar \psi_3 \psi_1 e^{-i {\pi\over 2} g},\\ \label{eq:44b}
\bar \Psi_4 \Psi_2 &= e^{+iQ_2}e^{+i{\pi\over 2} P_1} \bar \psi_4 \psi_2 e^{i {\pi\over 2} g},\\ \label{eq:44c}
\bar \Psi_3 \Psi_2 &= 
  e^{+iQ_1}e^{-i{\pi\over 2} P_2}  \bar \psi_3 \psi_2 e^{-i {\pi\over 2} g},\\ \label{eq:44d}
\bar \Psi_4 \Psi_1 &=e^{-iQ_1}e^{+i{\pi\over2}P_2} \bar \psi_4 \psi_1 e^{-i {\pi\over 2} g}.
\end{align}
\end{subequations}
%
We note that the zero mode sectors of the scattering operators 
in Eqs.~(\ref{eq:44a}), (\ref{eq:44b}) and Eqs.~(\ref{eq:44c}), (\ref{eq:44d}) 
 now pairwise commute among themselves. However, those in 
 Eqs.~(\ref{eq:44a}), (\ref{eq:44b}) and Eqs.~(\ref{eq:44c}), (\ref{eq:44d})
 do not commute. 
 This structure suggests a further simplification: from the commutation relation
\begin{align}
  \label{eq:44}
  [Q_i,P_j]=g i\delta_{ij}
\end{align}
we compute
\begin{align}
  \label{eq:46}
  [Q_2+{\pi\over 2} P_1,Q_1-{\pi\over 2} P_2]=-ig\pi.
\end{align}
Now this relation motivates the canonical transformation 
\begin{align}
  \label{eq:47}
  X&={Q_2\over \sqrt \pi} + {\sqrt \pi\over 2} P_1,\qquad
Y=-{Q_1\over \sqrt \pi} + {\sqrt \pi\over 2} P_2,\cr
  X'&={Q_1\over \sqrt \pi} + {\sqrt \pi\over 2} P_2,
\qquad
  Y'=-{Q_2\over \sqrt \pi} + {\sqrt \pi\over 2} P_1.
\end{align}
The new variables obey the relations $[X,Y]=ig$ and the same for
$X',Y'$. The primed and unprimed variables are mutually
commutative. All this means that the canonical piece of the action
remains invariant. In particular, we have a contribution
\begin{align}
  \label{eq:48}
  S[X,Y]={i\over g}\int d\tau\, Yd_t X \ .
\end{align}
The scattering operators assume the form
\begin{align}
  \label{eq:49}
  \bar \Psi_3 \Psi_1 &= e^{-i\sqrt \pi X}\bar \psi_3 \psi_1 e^{-i {\pi\over 2} g},\cr
\bar \Psi_4 \Psi_2 &= e^{i\sqrt \pi X} \bar \psi_4 \psi_2 e^{i {\pi\over 2} g},\cr
\bar \Psi_3 \Psi_2 &= 
e^{-i\sqrt \pi Y} \bar \psi_3 \psi_2 e^{-i {\pi\over 2} g},\cr
\bar \Psi_4 \Psi_1 &=e^{i\sqrt \pi Y}\bar \psi_4 \psi_1 e^{-i {\pi\over 2} g},
\end{align}
which means that we do not need  the variables $X',Y'$, and do not consider them in the following. 

Using the canonical representation \eqref{captialphi.eq} and assuming locking
of the charge mode we have
\begin{align}
  \label{eq:44}
  &\bar \Psi_3\Psi_1 \sim e^{i (-\sqrt \pi X + \sqrt 2
    \Phi_2)},\cr
&\bar \Psi_3\Psi_2 \sim e^{i(- \sqrt \pi Y - \sqrt 2
    \Phi_1)},
\end{align}
where we have absorbed  the $\exp(\pm i \sqrt \pi/2 g)$ factors  in
the scattering phases $\gamma_{13}$ and  $\gamma_{23}$.

The structure of the scattering operators suggests a shift
\begin{align}
  \label{eq:44}
  \Phi_2 &\to \Phi_2 + \sqrt{{\pi\over 2}}X,\cr
  \Phi_1 &\to \Phi_1 - \sqrt{{\pi\over 2}}Y,
\end{align}
whereupon the scattering operators simplify to 
\begin{align}
  \label{eq:44}
  &\bar \Psi_3\Psi_1 \sim e^{i \sqrt 2
    \Phi_2},\cr
&\bar \Psi_3\Psi_2 \sim e^{-i \sqrt 2
    \Phi_1},
\end{align}
As a result of this shift, the fields $\Phi_1$ and $\Phi_2$ no longer commute with each other, and we have 
\begin{align}
& [\Phi_2, \Phi_1] = i g {\pi \over 2}  \ \ . \label{Phicommutator.eq}
\end{align}

Now we are in a position to express the local effective action in terms of
variables which allow an efficient analysis of the competing scattering
processes. The dissipative part of the action, obtained by integrating out the
finite wave vector modes of the bosonic fields, is given by
%
\begin{align}
  \label{eq:50}
  S_{\rm diss}[\Phi]={T\over \pi g}\sum_{i=0}^2 \sum_n |\omega_n|\, |\Phi_{i,n}|^2 \ \ ,
\end{align}
%
where $\Phi_{i,n}$ denotes the n-th Matsubara component of the field $\Phi_i$.
The sum runs only up to $i=2$, on account of the locking of the field $\Phi_3$.
The $i=0$ contribution to the sum defines the zero mode action~\eqref{SPhi0},
and will be ignored throughout. Introducing the shorthand notation
\begin{align}
  \label{eq:51}
  \Phi=(\Phi_1,\Phi_2)^T,\qquad \Xi=(X,Y)^T,
\end{align}
the quadratic part of the action then assumes the form
%
\begin{eqnarray}
  S_0[X,\Phi] & = & {T\over \pi g}\sum_n \Phi^T_n \Phi_{-n} |\omega_n| -
  {T\over g} {2\over \pi} \sum_n \Phi^T_n \sigma_3 \Xi_{-n} |\omega_n|
  \nonumber \\  \label{eq:44}
 & &  + {T\over 2g}\sum_n \Xi^T_n
  \left(
    \begin{matrix}
      |\omega_n|&\omega_n\cr
      -\omega_n &|\omega_n|
    \end{matrix}
  \right)\Xi_{-n}.
\end{eqnarray}
%
The coupling between $\Phi$ and $\Xi$ makes sure that the commutation relation
Eq.~(\ref{Phicommutator.eq}) is reproduced in correlation functions of the
$\Phi_i$. Since the vector variable $\Xi$ no longer appears in the scattering
part of the action,  it can be integrated out. Doing the Gaussian integral, we
obtain the action $S_0[\Phi_1,\Phi_2]$ as given in Eq.~\eqref{eq:52}.

We finally note that the two
'vertical'  scattering operators $\sim \bar \Psi_3\Psi_1/\bar \Psi_4
\Psi_2$ (and similarly for  the horizontals) may couple at arbitrary
strength/scattering phase. This leads us to
$S[\Phi]=S_0[\Phi_1,\Phi_2]+S_v[\Phi_2]+S_h[\Phi_1]$, where the scattering action reads (cf. Eq.~\eqref{SScattPsi})
%
\begin{align} 
\label{eq:Svh}
&S_v[\Phi_2]=\sum_{\mathrm{x}=13,24}\int d\tau\,|\gamma_{\mathrm{x}}|\cos(\sqrt{2} \Phi_2+\phi_{\mathrm{x}}),\cr
&S_h[\Phi_1]=\sum_{\mathrm{x}=14,23}\int d\tau\,|\gamma_{\mathrm{x}}|\cos(\sqrt{2}\Phi_1+\phi_{\mathrm{x}}). 
\end{align}
%
Here, the phases $\phi_{ij}$  absorb the phase of the complex scattering amplitudes $\gamma_{ij}$, the phases 
$\exp(i\pi g/2)$ appearing in \eqref{eq:44}, and the relative sign change of
the field variables in $13$ vs $24$ (cf. Eq. \eqref{eq:49}).

Without loss of generality, we can use an addition theorem
for trigonometric functions followed by a constant shift of the fields
$\Phi_1$ and $\Phi_2$ to transform the scattering actions into the form given
in Eq.~\eqref{eq:52}, with $\gamma_v = | \gamma_{13} + e^{- i \pi g} \, \gamma_{24}|$ and 
$\gamma_h = | \gamma_{23} + \gamma_{14}|$.

\section{Non-Chiral Luttinger Liquids}

Throughout the analysis presented in this paper, we have assumed a setup consisting of a quantum dot connected to 4 chiral wires, which are geometrically separated from each other. These chiral wires ÒtalkÓ to each other through the tunneling to the QD (and the charging interaction thereon), and through elastic tunneling terms between the ÒincomingÓ and the ÒoutgoingÓ channels.  Such channels can be realized, for example, as the edges of a fractional quantum Hall strip; the most relevant tunneling operators are those of  fractionally charged quasiparticles (the latter possess fractional statistics as well), and we have assumed that such tunneling terms are allowed.  While intra-(chiral)channel interaction is allowed, inter-channels interaction is excluded. 

When it comes to realizing our theory with non-chiral Luttinger liquid wires, the situation is trickier. There are two main issues that should be noted. First, throughout our analysis, we have assumed that ÒforwardÓ and ÒbackwardÓ (elastic) scattering are treated on equal footing. There is no concrete significance  associated with ÒforwardÓ or ÒbackwardÓ. When it comes to non-chiral Luttinger wires, unless special conditions are specified, forward (elastic) scattering is irrelevant, and is clearly distinct from backward scattering. Second, as was noted above, in the case of chiral edges supported by an incompressible fractional quantum Hall electron gas,  it is clear that tunneling of quasi-particles is allowed. The situation with non-chiral Luttinger wires is different, however. In the analysis depicted below we allow only for electron tunneling; in general the results of the ensuing analysis will be qualitatively different. However, for a specific choice of the interaction (compare Eq.~(\ref{velocitymatrix.eq})), our beam splitter can be realized using non-chiral Luttinger liquid wires. 

Let us briefly review these two types of processes. We briefly repeat the analysis of Ref.~[\onlinecite{FisherGlazman}].  Consider the Luttinger liquid Hamiltonian 
%
\begin{equation}
H \ = \ {v \over 2 \pi} \big[ K (\partial_x \phi)^2 \ + \ K^{-1} (\partial_x \theta)^2 \big] \ \ , 
\end{equation}
%
where the bosonic fields satisfy a Kac-Moody algebra
%
\begin{equation}
\big[ \phi(x), \theta(x^\prime)\big] \ = \ {i \pi \over 2} \sgn (x - x^\prime) \ \ .
\end{equation}
%
Here, $K$ is the Luttinger liquid interaction parameter. One can define left and right moving chiral fields, $\Phi_{R/L} = \phi \pm \theta$, and respective electronic field operators $\psi_{R/L} = e^{\pm i k_F x } e^{i \Phi_{R/L}}$, with commutation relations $ \big[ \Phi_R(x), \Phi_R(x^\prime)\big] = - \big[ \Phi_L(x), \Phi_L(x^\prime) \big]\ = \ i \pi 
\sgn (x - x^\prime)$. In terms of these chrial modes, the Hamiltonian reads
%
\begin{equation}
H \ = \ \pi v_0 \big[ \rho_R^2 \ + \ \rho_L^2 \ + \ 2 \lambda \rho_R \rho_L \big] \ \ , 
\end{equation}
%
with an inter-channel interaction (between the two chiral modes). Here, 
%
\begin{equation}
\rho_{R/L} (x) \ = \ \pm {1 \over 2 \pi} \partial_x \Phi_{R/L}  \ \ .
\end{equation}
%
One can define other modes, $\varphi_{R/L} = K \phi \pm \theta$, in terms of which the Hamiltonian decouples into "left" and "right" sectors. These new modes have commutation relations 
$ \big[ \varphi_R(x), \varphi_R(x^\prime)\big] = - \big[ \varphi_L(x), \varphi_L(x^\prime) \big] \ = \ i \pi K\, 
\sgn (x - x^\prime)$. The operators $e^{i \varphi_{R/L}}$ are field operators of chiral quasi-particles that carry charge $K e$. However, experimentally realizable 
tunneling processes can only involve elctrons, and in the following we will discuss the RG relevance of such processes.

\subsection{Interactions within one LL only}

\begin{figure}[t]
\centering
\includegraphics[width=0.8\linewidth]{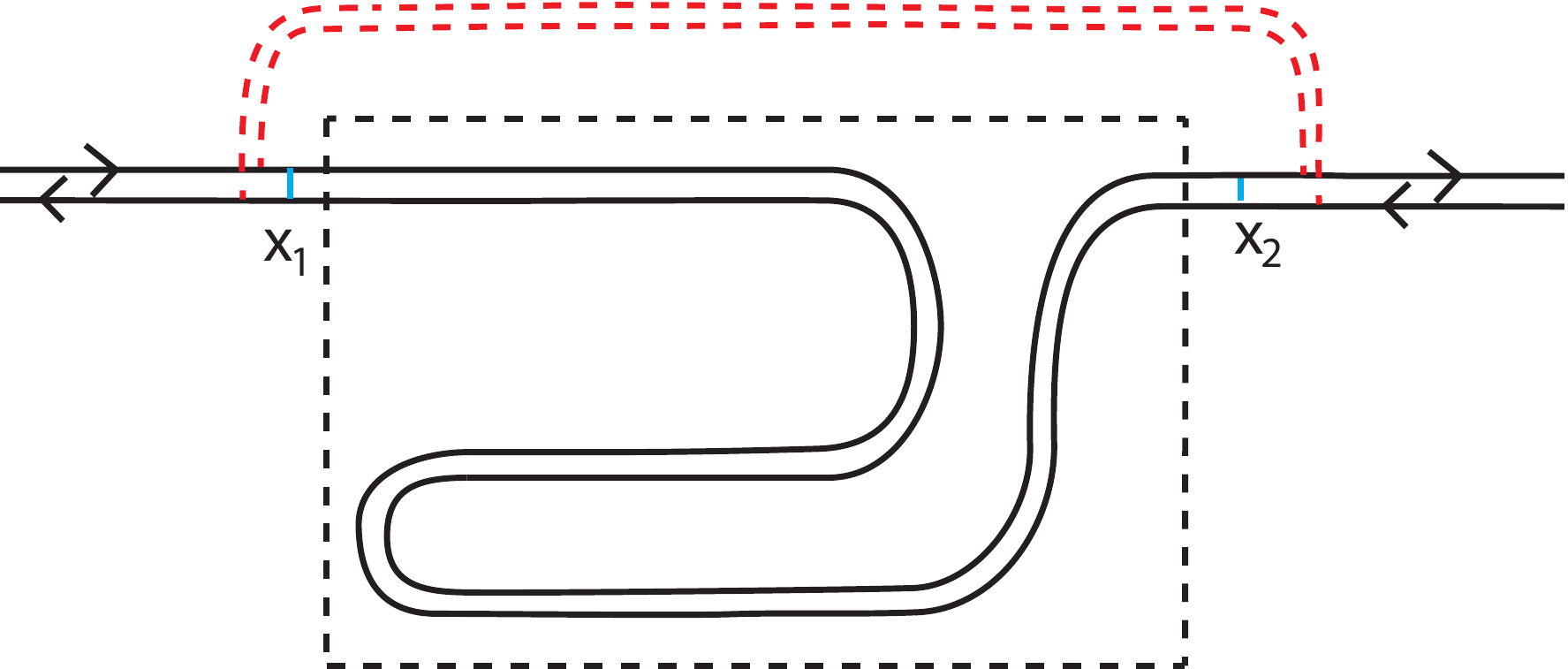}
\caption{\label{fig:nonchiral_local} It is not possible to realize a beam splitter with two standard non-chiral LLs because backscattering 
between left and right moving electrons in the same LL (short blue lines) is relevant, whereas forward scattering across the dot
(dashed red lines) is irrelevant.  The part of the wires inside the dashed box  constitutes the quantum dot with a nonlocal charging interaction. 
}
\end{figure}

A sketch of  a possible realization is shown in Fig.~\ref{fig:nonchiral_local}. It is well known that that a local potential (or, in the microscopic model, a weak bond) gives rise to backscattering and is a relevant perturbation in the RG sense. More problematic is the microscopic realization of an RG-relevant forward scattering term, which coherently transfers electrons across the quantum dot (dashed red lines in Fig.~1). In order to compute the scaling dimension of such a forward scattering term, we describe the LL by chiral bosonic eigenstates
$\varphi_\pm$ with imaginary time action 
%
\begin{equation}
S_{\rm L/R} = {u \over 4 \pi K} \int dx d\tau \,  \partial_x \varphi_{\rm R/L} \left(   \pm i \partial_\tau + 
\partial_x \right)\varphi_{\rm R/L}  \ \ . 
\end{equation}
%
Here, $u$ denotes the renormalized velocity, $K$ the LL parameter, and the smooth part of the total electron density is given by
$\rho(x) = {1 \over 2 \pi} (\partial_x \varphi_R + \partial_x \varphi_L)$. In this basis, right and left moving electrons (i.e. electronic states 
at the left and right Fermi point of the noninteracting system) have the bosonized form 
%
\begin{equation}
\psi_{\rm R/L}(x) = {1 \over \sqrt{2 \pi a} } e^{i[ K_{\pm} \varphi_R(x) - K_{\mp} \varphi_L(x)]} \ \ ,
\end{equation}
%
with $K_\pm = (K^{-1} \pm 1)/2$. Thus, a local backscattering operator is given by
%
\begin{eqnarray}
\hat{O}_{\rm }(x_1)& =& \psi_{\rm R}^\dagger(x_1) \psi_{\rm L}(x_1)\\
& =&  {1 \over 2 \pi a} e^{i(K_+ - K_-)[\varphi_{\rm R}(x_1) + \varphi_{\rm L}(x_1)]} \ \ . \nonumber
\end{eqnarray}
%
Using $K_+ - K_1 = 1$, we reproduce the well known result that the correlation function $\langle \hat{O}_{\rm back}(t) \hat{O}_{\rm back}(0)\rangle \sim t^{- 2 K}$, which makes backscattering a relevant perturbation for a repulsive interaction with $K < 1$.  On the other hand, a forward scattering term involves electron operators at two different spatial positions
%
\begin{eqnarray}
\hat{O}_{\rm f}(x_1,x_2)& =& \psi_{\rm R}^\dagger(x_2) \psi_{\rm R}(x_1) \label{forward.eq}\\
&=&  
{1 \over 2 \pi a} e^{i K_+ [\varphi_{\rm R}(x_1) - \varphi_{\rm R}(x_2)] - i K_- [ \varphi_{\rm L}(x_1) - \varphi_{\rm L}(x_2)]} \ \ .
\nonumber
\end{eqnarray}
%
Since positions $x_1$ and $x_2$ are separated by an "infinitely long" dot region, correlation functions between fields at positions 
$x_1$ and $x_2$ vanish. As a consequence, the correlation function in time of the forward scattering operator factorizes 
according to 
%
\begin{widetext}
\begin{eqnarray}
\langle \hat{O}_{\rm f}(t) \hat{O}_{\rm f}(0)\rangle & \propto &  e^{
- {1 \over 2} K_+^2 \langle[\varphi_{\rm R}(x_1,t) - \varphi_{\rm R}(x_1,0)]^2 \rangle 
- {1 \over 2} K_+^2 \langle[\varphi_{\rm R}(x_2,t) - \varphi_{\rm R}(x_2,0)]^2 \rangle } 
 e^{
- {1 \over 2} K_-^2 \langle[\varphi_{\rm L}(x_1,t) - \varphi_{\rm L}(x_1,0)]^2 \rangle  
- {1 \over 2} K_-^2 \langle[\varphi_{\rm L}(x_2,t) - \varphi_{\rm L}(x_2,0)]^2 \rangle
} 
\nonumber \\
& \propto & t^{- K( 2 K_+^2 + 2 K_-^2)}\nonumber \\
 & \propto & t^{- (K^{-1} + K)}  \ \ . 
\end{eqnarray}
\end{widetext}
%
Since the function $f(K) = K^{-1} + K$ has its  minimum value $2$ at $K=1$, the forward scattering operator Eq.~(\ref{forward.eq}) 
seems is irrelevant for all values of $K$. As a consequence, finding a microscopic description of an RG relevant forward scattering 
operator requires a modification of the interaction term.

\subsection{Local interactions between both nonchiral Luttinger liquids}

\begin{figure}[t]
\centering
\includegraphics[width=0.8\linewidth]{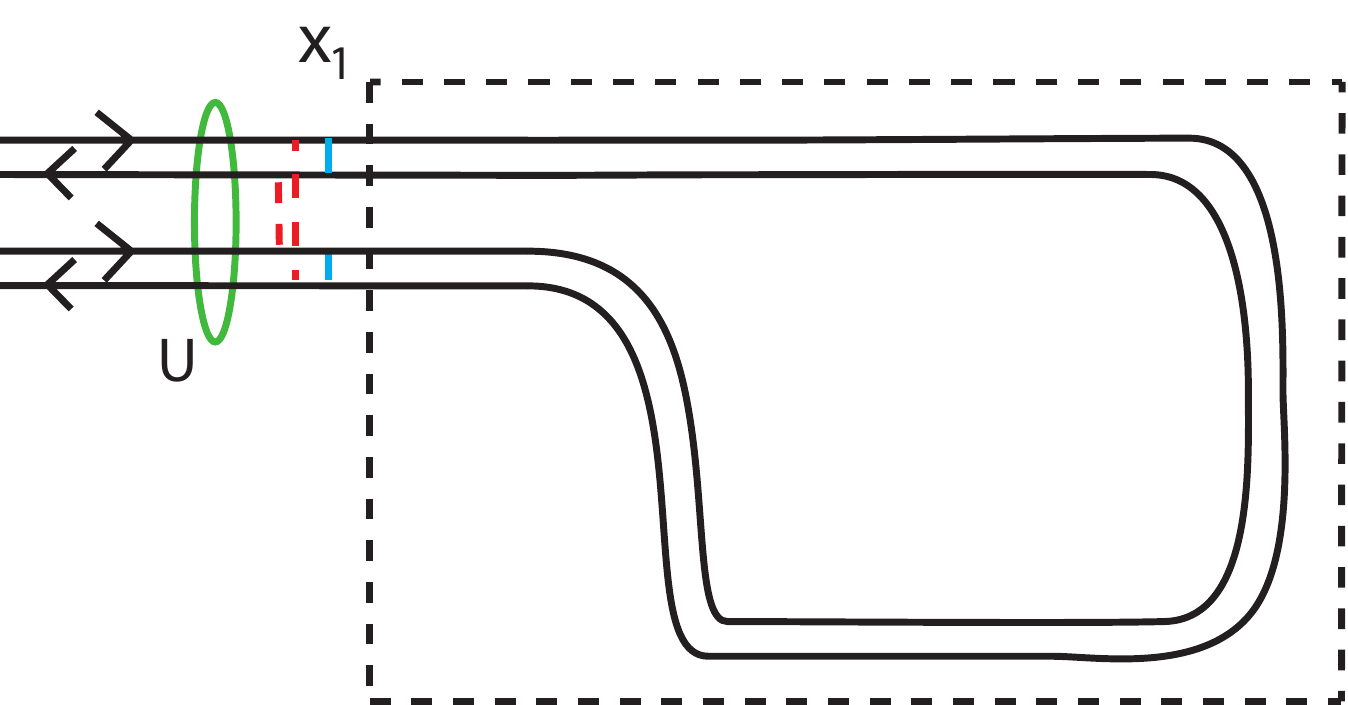}
\caption{\label{fig:nonlocal_interaction} In order to realize a beam splitter with a non-chiral setup, two LLs with a local charging interaction {\it which does not discriminate between the two wires} is needed (indicicated by green oval, strength $U >0$). In this setup, both intra-wire backscattering (short blue lines) and inter-wire backscattering (dashed red lines) are relevant, whereas inter-wire forward scattering (not shown in the figure) is irrelevant.  The part of the wires inside the dashed boxed constitutes the quantum dot with a nonlocal charging interaction. 
}
\end{figure}

Here, we want to allow for the possibility that there is local interaction term including both LL wires as shown in Fig.~\ref{fig:nonlocal_interaction}.
Following the notation of Wen (Quantum Field Theory of Many-Body systems, section 7.4.6), we describe the system by a $K$-matrix 
with $K = {\rm diag}(1,-1,1,-1)$, where we imagine that the first two branches belong to the first LL with fields $\varphi_{1R,0}$ and 
$\varphi_{1L,0}$, and the second two branches belonging to the second LL. These are eigenmodes in the absence of interactions. 
A local charging term acting on both LL wires, i.e.~equally on all four modes is  described by the velocity matrix
%
\begin{equation}
V = \left(
\begin{array}{cccc}  1 + U & U & U & U \\ U & 1+U & U & U \\ U & U & 1+U & U \\ U & U & U & 1+U
\end{array}
\right) \ \ . 
\label{velocitymatrix.eq}
\end{equation}
%
We now diagonalize the real symmetric velocity matrix by an orthogonal transformation according to $V \to V_1 = A V A^T$
with 
%
\begin{equation}
A = \left(
\begin{array}{cccc}  1/2 & 1/2 & 1/2 & 1/2 \\ 1/\sqrt{2} & - 1/\sqrt{2} & 0 & 0  \\ 0 & 0 & 1/\sqrt{2} & - 1/\sqrt{2}    \\ 1/2 & 1/2 & -1/2 & 
- 1/2 
\end{array}
\right) \ \ . 
\end{equation}
%
We obtain $V_1 = {\rm diag}( 1 + 4 U,1,1,1)$. In the following, we introduce the charge mode velocity 
%
\begin{equation}
u \ = \ 1 + 4 U \ \ . 
\end{equation}
%
Next, we rescale fields such that 
the velocity matrix becomes equal to the unit matrix, i.e. $V_2 = \Lambda V_1 \Lambda$ with $\Lambda = {\rm diag}(1/\sqrt{u},1,1,1)$. With the help of these transformations, we find $K_2 = \Lambda A K A^T \Lambda$ with 
%
\begin{equation}
K_2 = \left(
\begin{array}{cccc}  0 & 1/\sqrt{2 u} & 1/\sqrt{2 u}& 0  \\ 1/\sqrt{2 u} & 0 & 0 & 1/\sqrt{2}    \\ 1/\sqrt{2 u} & 0 & 0 & -1/\sqrt{2}    \\
0 & 1/\sqrt{2} & - 1/\sqrt{2} & 0
 \end{array}
\right) \ \ . 
\end{equation}
%
Since $K_2$ is real symmetric, it can be diagonalized by an orthogal transformation according to $K_d = B K_2 B^T$ with 
%
\begin{equation}
B= \left(
\begin{array}{cccc}   - 1/\sqrt{2} & 1/2 & 1/2 & 0 \\  1/\sqrt{2} & 1/2 & 1/2 & 0    \\    0 & - 1/2 & 1/2 & 1/\sqrt{2}    \\ 0 & 1/2 & - 1/2 & 1/\sqrt{2}
\end{array}
\right) \ \ ,
\end{equation}
%
and we obtain $K_d = { \rm diag}(- 1/\sqrt{u},1/\sqrt{u},-1,1)$. The entries of $K_d$ are the inverse of the scaling dimensions of the new fields. 
Through this series of transformations, the new fields $\phi_1, \phi_2, \phi_3 ,\phi_4$ are related to the original fields according to 
the transformation $\phi = B \Lambda^{-1} A \varphi$. In order to compute scaling dimensions of tunneling operators, we need the inverse transformation which is given by 
%
\begin{equation}
C \equiv ( B \Lambda A)^{-1} = A^T \Lambda B^T =  \left(
\begin{array}{cccc} {-1 + \sqrt{u} \over 2 \sqrt{2 u}}   &  {1 + \sqrt{u} \over 2 \sqrt{2 u}}  & 0 & {1\over \sqrt{2}}   \\[.3cm]  
- {1 + \sqrt{u} \over 2 \sqrt{2 u}} & {1 - \sqrt{u} \over 2 \sqrt{2 u} } & {1 \over \sqrt{2}} & 0       \\[.3cm]
  {-1 + \sqrt{u} \over 2 \sqrt{2 u}}   &  {1 + \sqrt{u} \over 2 \sqrt{2 u}}  & 0 & -{1\over \sqrt{2}}\\[.3cm]
  - {1 + \sqrt{u} \over 2 \sqrt{2 u}} & {1 - \sqrt{u} \over 2 \sqrt{2 u} } & -{1\over \sqrt{2}} & 0 
\end{array}
\right) \ \ .
\end{equation}
%
Using the matrix $C$, we can express the original fields in terms of the new ones according to 
%
\begin{equation} 
\varphi = C \phi \ \ . 
\end{equation}
%
In order to compute the scaling dimension of the operator for backscattering a right mover in wire one into a left mover in wire two, 
we use the expression 
%
\begin{eqnarray}
\hspace*{-.8cm}
\varphi_{1R} + \varphi_{2L}  & = & - {1 \over \sqrt{2 u}} \phi_1+ { 1 \over \sqrt{2 u}} \phi_2 - {1 \over \sqrt{2}} \phi_3  +  
\phi_4   ,
\end{eqnarray}
%
and find for this tunneling operator the  scaling dimension 
%
\begin{eqnarray}
g_{1R,2L} & = & {1 \over 2} \left(  \sqrt{u} \, {1 \over 2 u} \ + \ \sqrt{u} \, {1 \over 2 u} \ + \ {1 \over 2} \ + \ {1 \over 2} \right) \nonumber \\
& = & {1 \over 2} \left( {1 \over \sqrt{u} } \ + \ 1 \right) \ \ .
\end{eqnarray}
%
In the presence of a repulsive interaction $U > 0$ the charge mode velocity  $u >1$, and hence $ g_{1R,2L} < 1$ makes interwire backscattering relevant. Similarly, we find that intra-wire backscattering is relevant with $g_{1R,1L} =  g_{1R,2L}$. Thus, this model allows to realize the intermediate fixed point discussed in the main text in the framework of non-chrial LL wires.

\section{Solution of the refermionized model for $g=1$}
\label{sec:refermionization}

In this appendix, we discuss the derivation of the scattering probabilities Eqs.~(\ref{P_14.eq}), (\ref{P_24.eq}).

\subsection{Solution for a single scattering amplitude}

We first consider the simple situation where $\gamma_{v}=0$ and only $\gamma_{h} \neq 0$.  In order to construct scattering states as eigenstates of the Hamiltonian, we need to make assumptions about the commutators of $\eta_1 \eta_4$ with $f_1$ and $\Psi_1$. If $\eta_1 \eta_4$ commutes with both $f_1$ and $\Psi_1$, we will arrive at the standard solution of refermionization \cite{Chamon+96}. As there is no reason to expect that our result should differ from the standard one, we will make this assumption. Then, we define the new Majorana operator
$\tilde{f}_1 = i \eta_1 \eta_4 f_1$, and with $H = H_0 + H_{\rm tun}$ find the following equations of motion 
%
\begin{subequations}
\begin{eqnarray}
- i \partial_t  \Psi_1(x) & = &  [ H, \Psi_1(x)]  \label{motion_gammahonly_a.eq}\\
& =  &   i \partial_x \Psi_1(x) \ + \ \sqrt{2 \pi a}\ \gamma_{h}
\ \tilde{f}_1 \ \delta(x) \nonumber \\[.5cm]
- i \partial_t  \Psi^\dagger_1(x) & = &  [ H, \Psi^\dagger_1(x)] \label{motion_gammahonly_b.eq}\\
  & = &   i \partial_x \Psi^\dagger_1(x) \ - \ \sqrt{2 \pi a}\ \gamma_{h} \ \tilde{f}_1 \ \delta(x) \nonumber \\[.5cm]
- i \partial_t \tilde{f}_1 & = & [ H, \tilde{f}_1] \label{motion_gammahonly_c.eq}\\
& =  &  2 \sqrt{2 \pi a} \ \gamma_{h}\left[  \Psi_1(0)  -  \Psi^\dagger_1(0) 
\right] \ \ . \nonumber
\end{eqnarray}
\end{subequations}
%
Away from $x=0$, $\Psi_1(x)$ satisfies the free fermion equation and has plane waves as a solution. At $x=0$, the field 
is discontinuous and acquires a phase shift. As a consequence, in the above equations of motion, $\Psi(0)$ needs to be interpreted as ${1\over 2} [\Psi_1(0+) + \Psi_1(0-)]$.   We solve the equations of motions by using the most general expression for  scattering eigenstates
%
\begin{eqnarray}
A_{1, \epsilon}^\dagger & = & \int_x \left(  \varphi_{11,\epsilon} \Psi_1^\dagger + \varphi_{14,\epsilon}^*  \Psi_1   + a_{11,\epsilon} \tilde{f}_1 \right)
\end{eqnarray}
%
For the wave functions, we make the ansatz
%
\begin{subequations}
\begin{eqnarray}
\varphi_{11,\epsilon}(x)  & = & \theta(-x) e^{i \epsilon x} \ + \ \theta(x) e^{i \delta_{11} } e^{i \epsilon x} \\
\varphi^*_{14,\epsilon}(x)  & = & \theta(x) e^{i \delta_{14} } e^{i \epsilon x} \ \ .
\label{ansatz_gammahonly.eq}
\end{eqnarray}
\end{subequations}
%
Following Ref.~\onlinecite{Chamon+96}, we interpret the scattering states in the following way: the region $x<0$ corresponds to the incoming wave packet, 
the region $x > 0 $ the scattered components of the wave packet. The wave function $\varphi_{11}$ multiplying the creation operator $\psi^\dagger$ describes the amplitude for forward scattering, i.e.~is associated with the component of the scattering state which continues to 
propagate along segment $1$ of the edge, in the interior of the QD. The wave function $\varphi_{14}^*$ multiplying the annihilation operator $\psi_1$ on the other hand 
is interpreted as the amplitude for scattering onto the outgoing part of edge segment $4$. Clearly, unitarity demands that the sum of the probabilities for forward and backwards scattering is equal to one, $|\varphi_{11}|^2 + |\varphi_{14}|^2 = 1$.  

Using the ansatz Eq.~(\ref{ansatz_gammahonly.eq}) in the equations of motion Eqs.~(\ref{motion_gammahonly_a.eq})-(\ref{motion_gammahonly_c.eq}), we find 
%
\begin{subequations}
\begin{eqnarray}
& & i\left[1 - e^{i \delta_{11}}\right] - 2 \sqrt{2 \pi a} \gamma_h a_{11}  =  0  \\
& & - i e^{i \delta_{14}} + 2 \sqrt{2 \pi a} \gamma_h a_{11} =0 \\
& & \epsilon a_{11}  =  {\sqrt{2 \pi a} \over 2} \gamma_h \left[ - \left( 1 + e^{i \delta_{11}}\right) + e^{i \delta_{14}} \right] \ \ , 
\end{eqnarray}
\end{subequations}
%
and finally the solutions 
%
\begin{subequations}
\begin{eqnarray}
e^{i \delta_{11}} & = & {i \epsilon \over i \epsilon - 4 \pi a \gamma_h^2} \\[.5cm]
e^{i \delta_{14}} & = & { - 4 \pi a \gamma_h^2 \over i \epsilon - 4 \pi a \gamma_h^2}  \ \ .
\end{eqnarray}
\end{subequations}
%
We note that, different from the case of truely non-interacting fermions, $|\varphi_{11,\epsilon}|^2$ and $|\varphi_{14,\epsilon}|^2$ cannot be interpreted as 
scattering probabilities for each value of energy separately. Instead, in order to obtain the scale dependent scattering probability, they need to be integrated over the derivative of the Fermi function to obtain
%
\begin{eqnarray}
P_{14} & = & \int d \omega (- f^\prime(\omega)) |\varphi_{14,\omega}|^2 \nonumber \\
& \approx &  {2 \Gamma_h \over T}\ \arctan{T \over 2   \Gamma_h} \ \ .
\end{eqnarray}
%
In the last step, the derivative of the Fermi function was approximated as  a box function of width temperature $T$ and height $1/T$.
Identifying the width of the resonance as $\Gamma_h = 4 \pi a \gamma_h^2$, and introducing the scaling function
$ F(x) = {2 \over x} \arctan{x\over 2}$,  this result can be written in a more compact form as $P_{14} = F(T/\Gamma)$. The scaling function has the limiting behaviors $F(x) = 1 - {1 \over 3} (x / 2)^2$ for small $x$ and $ F(x) = {2 \over x}$ 
for large $x$. Using these asymptotics, we find 
%
\begin{eqnarray}
P_{14} & = & \left\{ 
\begin{array}{cc}   {2 \Gamma_h \over T} & {\rm for } \ T \gg  \Gamma_h \\[.5cm]
1 - {T^2 \over 12 \Gamma_h^2} & {\rm for} \ T \ll \Gamma_h 
\end{array} \right.
\end{eqnarray}
%
This result agrees with the  standard perturbative analysis of a single scatterer in a LL. 

We now discuss a complementary scenario, in which $\gamma_h= 0 $ and $\gamma_v \neq 0$. By symmetry, it is clear that the solution can be found in an analogous way as discussed above. For future reference, the solution for the scattering eigenstate is given by 
%
\begin{eqnarray}
A_{2, \epsilon}^\dagger & = & \int_x \left(  \varphi_{22,\epsilon} \Psi_2^\dagger + \varphi_{44,\epsilon}^*  \Psi_2   +
 a_{22,\epsilon} \tilde{f}_1 \right) \ \ .
\end{eqnarray}
%
Now, for $x < 0$ the operator $\psi_2^\dagger$ describes an incoming state along countour $2$. For $x > 0$, the operator $\psi_2^\dagger$ creates a state which is forward scattered into the interior of the QD along contour $2$, whereas $\psi_2$ describes a partial wave which is scattered 
onto the outgoing part of edge $4$. In what follows, it will be important that both $\psi_1(x)$ and $\psi_2(x)$ create an outgoing wave on 
edge $4$.

\subsection{Solution for two competing scattering amplitudes}

We now consider the interesting case in which both $\gamma_h$ and $\gamma_v$ are nonzero. Due to the Klein factors represented 
by products of Majorana fermions $\eta_1 \eta_4 $ and $\eta_2 \eta_4$, the two different scattering operators do not commute with each other, and the 
respective equations of motion are coupled. In addition to $\tilde{f}_1 = i \eta_1 \eta_4 f_1$, we now introduce $\tilde{f}_2 = i \eta_2 \eta_4 f_1$.  Now we  need to take into account the nontrivial anti-commutator
%
\begin{equation}
\{ \tilde{f}_1, \tilde{f}_2 \}\  = \ - 2 \eta_1 f_1 \eta_2  f_2  \ \equiv \ -2 \,  \hat{C}  \ \ ,
\end{equation}
%
where we have made the assumption that $f_1$ and $f_2$ commute with each other and with $\eta_1\eta_4$ and $\eta_2\eta_4$. 
Since $\hat{C}$  is a product of an even number of Majoranas, it trivially commutes with $\Psi_1$, $\Psi_1^\dagger$, $\Psi_2$, and 
$\Psi_2^\dagger$. In addition, 
%
\begin{equation}
[ \tilde{f}_1, \hat{C}] \ = \ 0 \ \ \ \ ,  \ {\rm and} \ \ \ \ \  [\tilde{f}_2 , \hat{C} ] = 0  \ \ \ \ \Rightarrow   \ \  [H_{\rm tun}, \hat{C} ] =0 \ \ . 
\end{equation}
 %
 As $\hat{C}$ commutes with all operators, it commutes with the Hamiltonian and hence  does not have any dynamics. In addition, it squares to one $\hat{C}^2=1$. With this in mind, we find the following equations of motion for the full system:
 %
 \begin{widetext}
 \begin{subequations}
 \begin{eqnarray}
  - i \partial_t  \Psi_1(x) & = &  [ H, \Psi_1(x)]  \ = \  i \partial_x \Psi_1(x) \ + \ \sqrt{2 \pi a}\ \gamma_{h} 
\ \tilde{f}_1 \ \delta(x)\\[.5cm]
- i \partial_t  \Psi^\dagger_1(x) & = &  [ H, \Psi^\dagger_1(x)]  \ = \  i \partial_x \Psi^\dagger_1(x) \ - \ \sqrt{2 \pi a}\ \gamma_{h} \ \tilde{f}_1 \ \delta(x)\\[.5cm]
- i \partial_t \tilde{f}_1 & = & [ H, \tilde{f}_1] \ = \ 2 \sqrt{2 \pi a} \ \left[ \gamma_{h} \Psi_1(0)  - \gamma_{h}
 \Psi^\dagger_1(0)  - \hat{C}\,  \gamma_{v}  \Psi_2(0) \ + \ \hat{C}\, \gamma_{v} \Psi_2^\dagger(0)  
\right] \\[.5cm]
- i \partial_t  \Psi_2(x) & = &  [ H, \Psi_2(x)]  \ = \  i \partial_x \Psi_2(x) \ + \ \sqrt{2 \pi a}\ \gamma_{v} 
\ \tilde{f}_2 \ \delta(x)\\[.5cm]
- i \partial_t  \Psi^\dagger_2(x) & = &  [ H, \Psi^\dagger_2(x)]  \ = \  i \partial_x \Psi^\dagger_2(x) \ - \ \sqrt{2 \pi a}\ \gamma_{v} \ \tilde{f}_2 \ \delta(x)\\[.5cm]
- i \partial_t \tilde{f}_2 & = & [ H, \tilde{f}_2] \ = \ 2 \sqrt{2 \pi a} \ \left[ \gamma_{v} \Psi_2(0)  - \gamma_{v} \Psi^\dagger_2(0)  \ - \ \hat{C} \, \gamma_{h} \Psi_1(0) \ + \ \hat{C} \, \gamma_{h}^* \Psi^\dagger(0)
\right]
\end{eqnarray}
\end{subequations}
\end{widetext}
%
In the following, we will derive scattering states as a solution of these equations by using the fact that  $\hat{C}^2 = 1$.  
We make the most general ansatz for operators 
%
\begin{subequations}
\begin{eqnarray}
A_{1, \epsilon}^\dagger & = & \int_x \left(  \varphi_{11,\epsilon} \Psi_1^\dagger + \varphi_{14,\epsilon}^*  \Psi_1 +
\varphi_{12, \epsilon} \Psi_2^\dagger + \tilde{\varphi}^*_{14,\epsilon} \Psi_2\right. \nonumber \\
& & \left.  + a_{11,\epsilon} \tilde{f}_1 + a_{12,\epsilon} \tilde{f}_2
\right) \label{A1.eq} \\[.5cm]
A_{2, \epsilon}^\dagger & = & \int_x \left(  \varphi_{22,\epsilon} \Psi_2^\dagger + \varphi_{24,\epsilon}^*  \Psi_2 +
\varphi_{21, \epsilon} \Psi_1^\dagger + \tilde{\varphi}^*_{24,\epsilon} \Psi_1\right. \nonumber \\
& & \left.  + a_{22,\epsilon} \tilde{f}_2 + a_{21,\epsilon} \tilde{f}_1
\right) \ \ .  \label{A2.eq}
\end{eqnarray}
\end{subequations}
%
creating  eigenstates of the Hamiltonian $H$ and satisfying
%
\begin{equation}
 \epsilon   A_{1, \epsilon}^\dagger = [ H, A_{1, \epsilon}^\dagger] , \ \ \ \ 
\epsilon A_{2,  \epsilon}^\dagger  = [H, A_{2,  \epsilon}^\dagger] \ \ .
\label{eigenvalue.eq}
\end{equation}
%
In the following, we only discuss the expression for $A_{1, \epsilon}^\dagger$, as the corresponding expression for $A_{2,  \epsilon}^\dagger$ is obtained by interchanging $\gamma_{h}$ with 
$\gamma_{v}$. Specifically, we make the ansatz
%
\begin{subequations}
\begin{eqnarray}
\varphi_{11,\epsilon}(x)  & = & \theta(-x) e^{i \epsilon x} \ + \ \theta(x) e^{i \delta_{11} } e^{i \epsilon x} \\
\varphi^*_{14,\epsilon}(x)  & = & \theta(x) e^{i \delta_{14} } e^{i \epsilon x}\\
\varphi_{12,\epsilon} (x) & = & \theta(x)  e^{i \delta_{12}} e^{i \epsilon x}\\
\tilde{\varphi}^*_{14,\epsilon}(x)  & = & \theta(x) e^{i \tilde{\delta}_{14}} e^{i \epsilon x}  
\end{eqnarray}
\end{subequations}
%
Imposing the condition Eq.~(\ref{eigenvalue.eq}), the equations
%
\begin{subequations}
\begin{eqnarray}
& & \hspace*{-1cm} i \left[ 1 - e^{i \delta_{11}} \right] -   2 \sqrt{2 \pi a}\, \gamma_{h} \left[ a_{11,\epsilon}  - \hat{C} a_{12,\epsilon}  \right]  = 0  \\[.5cm]
& & \hspace*{-1cm}  -i e^{i \delta_{14}}  +  2 \sqrt{2 \pi a}\, \gamma_{h} \left[ a_{11,\epsilon} - \hat{C} a_{12,\epsilon} \right] = 0 \\[.5cm]
  & &\hspace*{-1cm} - i e^{i \delta_{12}}  -   2 \sqrt{2 \pi a} \, \gamma_{v} \left[ a_{12,\epsilon} - \hat{C} a_{11,\epsilon} \right]  =  0 \\[.5cm] 
 & & \hspace*{-1cm}-i e^{i \tilde{\delta}_{14}}  +  2 \sqrt{2 \pi a} \, \gamma_{v} \left[ a_{12,\epsilon} -  \hat{C}  a_{11,\epsilon}  \right]  =  0  \\[.5cm] 
 & &\hspace*{-1cm} \epsilon \, a_{11,\epsilon}  =  {\sqrt{2 \pi a} \over 2} \left[ - \left( 1 + e^{i \delta_{11}} \right) \gamma_{h} + e^{i \delta_{14} }
\gamma_{h} \right]\label{a11.eq}   \\[.5cm]
  & & \hspace*{-1cm} \epsilon \, a_{12,\epsilon} =  {\sqrt{2 \pi a} \over 2} \left[ - e^{i \delta_{12}} \gamma_{v} + e^{i \tilde{\delta}_{14}} \gamma_{v} \right]   \label{a12.eq}
\end{eqnarray}
\end{subequations}
%
are found. Using $\hat{C}^2 = 1$, we see that the first four equations only depend on the combination $a_{11,\epsilon}  - \hat{C} a_{12,\epsilon}$, and if we take an appropriate linear combination of Eqs.~(\ref{a11.eq}, \ref{a12.eq}), we can solve for 
%
\begin{subequations}
\begin{eqnarray}
e^{i \delta_{11}} & = & { i \epsilon \ - \ 4 \pi a \gamma_{h}^2 \over i \epsilon - 4 \pi a \left( \gamma_{h}^2   + \gamma_{v}^2 \right) } \\[.5cm]
e^{i \delta_{14}} & = & { - 4 \pi a \gamma_h^2 \over i \epsilon - 4 \pi a (\gamma_h^2 + \gamma_v^2)} \\[.5cm]
e^{i\delta_{12}} & = & {- 4 \pi a \gamma_h \gamma_v \over i \epsilon - 4 \pi a (\gamma_h^2 + \gamma_v^2)} \\[.5cm]
e^{i \tilde{\delta}_{14}} & = & { 4 \pi a \gamma_h \gamma_v \over i \epsilon - 4 \pi a (\gamma_h^2 + \gamma_v^2)}
\end{eqnarray}
\end{subequations}
%
The probability for forward scattering is given by $|e^{i \delta_{11}} |^2$, the probability for scattering onto lead $2$ is given by $|e^{i\delta_{12}}|^2$, and the probability for scattering onto lead $4$ is given by $|e^{i \delta_{14}}|^2 + |e^{i \tilde{\delta}_{14}} |^2$. There are two contributions for scattering onto lead $4$, one due to the operator $\psi_1$ in the scattering state Eq.~(\ref{A1.eq}), and a second one due to the operator $\psi_2$ 
in the same scattering state. We will see in a moment that indeed both these contributions are needed in order to reproduce the perturbative result 
for the tunneling probability from lead $1$ onto lead $4$. 

Integrating over the derivative of the Fermi function as in the case of a single scatterer, we obtain the probabilities for transmission 
%
\begin{subequations}
\begin{eqnarray}
P_{11} & = &  {\Gamma_v^2 \over \Gamma^2}\ F\left({T \over \Gamma}\right)\\
P_{14} & = &  {\Gamma_h \over \Gamma}\ F\left({T \over \Gamma}\right) \\
P_{12} & = & {\Gamma_h \Gamma_v \over \Gamma^2}\ F\left({T \over \Gamma}\right)
\end{eqnarray}
\end{subequations}
%
We now argue that these results agree with a perturbative calculation in $\gamma_h$ and $\gamma_v$. In the limit of large $T/\Gamma$, 
one finds  $P_{14} = 2 \Gamma_h/T$, exactly the same result as in the case $\gamma_v=0$ discussed previously. This is to be exptected, 
since corrections due to $\gamma_v$ can enter only in an additive fashion, and have to be of order $O(\gamma_h^2 \gamma_v^2)$. 
On the other hand, in the same limit, one finds $P_{12} = 2 \Gamma_h \Gamma_v /\Gamma T$, which is again to be expected since scattering from lead $1$ into lead $2$ has to be a two-step process and has to be proportional to $\gamma_h^2 \gamma_v^2$.




\end{document}